\documentclass[%
 reprint,
 amsmath,amssymb,
 aps,
prb,
]{revtex4-2}

\usepackage{graphicx}
\usepackage{dcolumn}
\usepackage{bm}

\begin{document}

\preprint{APS/123-QED}

\title{Probing the pairing symmetry in kagome superconductors based on the single-particle spectrum}

\author{Jinhui Liu}
\author{Tao Zhou}%
\email{Corresponding author: tzhou@scnu.edu.cn}
\affiliation{%
	Guangdong Basic Research Center of Excellence for Structure and Fundamental Interactions of Matter, Guangdong Provincial Key Laboratory of Quantum Engineering and Quantum Materials, School of Physics, Guangdong-Hong Kong Joint Laboratory of Quantum Matter, and Frontier Research Institute for Physics, South China Normal University, Guangzhou 510006, China
}%

\date{\today}

\begin{abstract}

We investigate the single-particle spectra of recently discovered kagome superconductors. We examine nine distinct superconducting pairing symmetries, including the uniform $s$-wave pairing function, nearest-neighbor and next-nearest-neighbor $s$-wave pairing, $p+ip$ pairing, $d+id$ pairing, and $f$-wave pairing states. These pairing states can be classified into irreducible representations of the $C_{6v}$ point group. Our findings suggest that these pairing states can be differentiated by analyzing the energy bands, spectral function, and local density of states. Additionally, we explore the single impurity effect, which could potentially assist in further distinguishing these pairing symmetries.
\end{abstract}

\maketitle


\section{\label{sec1.intr}INTRODUCTION}

The superconductivity has been discovered in several families of kagome lattice, including the Mg$_2$Ir$_3$Si~\cite{doi:10.7566/JPSJ.89.013701}, the RT$_3$X$_2$ family (R=Ce, La, T=4d or 5d transition metal, and X=Si, B or Ga)~\cite{KU198091,ATHREYA1985330,PhysRevB.84.214527,LI2015248,PhysRevB.107.085103,Gui2022,Gong_2022}, the vanadium-based AV$_3$Sb$_5$ (A=K, Rb, Cs) family~\cite{r11,r12,r13}, the titanium-based ATi$_3$Bi$_5$ (A=Rb, Cs)  family~\cite{yang2022titaniumbased,yang2022superconductivity,Li_2023}, the Ta$_2$V$_{3.1}$Si$_{0.9}$ material~\cite{liu2023vanadiumbased} and the CsCr$_3$Sb$_5$ material~\cite{xu2023frustrated,liu2023superconductivity}. 
In the past several years, the AV$_3$Sb$_5$ family has sparked significant interest. Several unique properties have been exhibited, such as the double superconducting dome feature in the temperature-pressure phase diagram~\cite{r14,r15,r16,r17,r18}, the time-reversal symmetry breaking charge order~\cite{r14,r15,r16,r17,r18}, the quantum anomalous Hall effect~\cite{doi:10.1126/sciadv.abb6003}, the lack of long-range magnetic ordering~\cite{r5}, a Z$_2$ topological band structure~\cite{r11}, and the possible Majorana zero modes~\cite{PhysRevB.106.085420}. Thus, this family presents a new platform for investigating the interplay between the unconventional superconductivity, competing orders, the lattice geometry, and the band topology. Furthermore, two recently reported kagome superconductors, Ta$_2$V$_{3.1}$Si$_{0.9}$~\cite{liu2023vanadiumbased} and CsCr$_3$Sb$_5$~\cite{xu2023frustrated,liu2023superconductivity}, exhibit a relatively higher superconducting transition temperature and strong correlation, respectively, providing new opportunities for studying superconductivity within the kagome lattice.

Identifying the pairing symmetry is an essential step for understanding the origin of the superconductivity. 
Theoretically, various possible pairing symmetries have been proposed by different groups, including the $s$-wave pairing symmetry~\cite{r31,r4,r30,PhysRevB.106.014501}, the $p$-wave pairing symmetry~\cite{r30}, the chiral $d+id$ pairing symmetry~\cite{r4,r2,r27,r28,r29}, and the $f$-wave pairing symmetry~\cite{r3,r29}. It was also proposed that the vortex and edge spectra can be used to diagnosis the pairing symmetries~\cite{PhysRevB.105.174518}. 
Experimentally, a variety of results have been proposed, primarily for the AV$_3$Sb$_5$ family.
Signature of nodal superconducting gap was revealed through the thermal conductivity measurements at the zero magnetic field~\cite{r15}. The existence of residual zero energy states and the $V$-shaped superconducting gaps from the scanning
tunneling microscopy (STM) experiment also seem to propose the existing of nodal points.
However, the measurements of the spin-lattice relaxation rate, the magnetic penetration depth, the impurity effect, and the transport measurements indicate that the superconducting AV$_3$Sb$_5$ 
is fully gapped~\cite{r19,r20,r24,r25,r26,r23}. Very recently, a nearly isotropic superconducting gap around the Fermi surface was revealed by the angle-resolved
photoemission spectroscopy (ARPES) experiment, providing substantial evidence of a fully gapped superconducting state~\cite{Zhong2023}. It was also reminded in Ref.~\cite{Zhong2023} that the pairing symmetry still cannot be confirmed, as ARPES experiments do not provide phase information of the superconducting gap.
Moreover, in recent years, there has been a continuous discovery of new kagome-lattice based superconducting materials with some unusual physical properties.
Therefore, now it is of importance and timely to investigate this issue theoretically and provide more useful information.

Theoretically, the ARPES and STM experimental results correspond to the momentum space and real space single-particle spectra, respectively, which are calculated from the imaginary parts of the Green's function. A useful link for
theoretical calculations and experimental observations can
be established. Moreover, the impurity effect may be theoretically investigated based on the single-particle spectrum and the $T$-matrix method~\cite{RevModPhys.78.373}.  There are also widespread proposals of in-gap bound states induced by the impurity to resolve pairing symmetries of various unconventional superconductors~\cite{RevModPhys.78.373,PhysRevLett.72.1526,PhysRevLett.103.186402,PhysRevB.80.064513,Li2021}. 
Thus, the investigation of single-particle spectra theoretically is of great importance and could provide useful information to probe pairing symmetry.

In this paper, we present a theoretical study of the single-particle spectra of a kagome superconductor. We primarily concentrate on the pure superconducting state, specifically in the phase diagram where the charge order is absent~\cite{r14,r15,r16,r17,r18}. 
Furthermore, we take into account all relevant pairing possibilities based on the based on the symmetry classification~\cite{PhysRevB.51.16233,RevModPhys.72.969,PhysRevB.90.054521,PhysRevB.92.085121,PhysRevB.100.214507}.
Nine possible pairing symmetries are considered to explore single-particle spectra, and the single impurity effect is also examined by calculating the local density of states (LDOS). It is suggested that the in-gap resonant states are induced near the impurity for $p+ip$ and $f$-wave pairing symmetries. We suggest that single-particle spectra can indeed be used to distinguish between different pairing symmetries.

The rest of the paper is organized as follows. In  Sec.~\ref{sec2.maf}, we introduce the model and present the relevant formalism. In Sec.~\ref{sec3.rad}, we report numerical calculations and discuss the obtained results. Finally, we present a brief summary in Sec.~\ref{sec4.sum}. 

\section{\label{sec2.maf}MODEL AND FORMALISM}
We start with a model including the normal state term ($H_N$) and the superconducting pairing term ($H_{SC}$), as expressed below,
\begin{equation}
	H=H_N+H_{SC}.
	\label{eq1}
\end{equation}

\begin{figure}
	\includegraphics[scale=0.25]{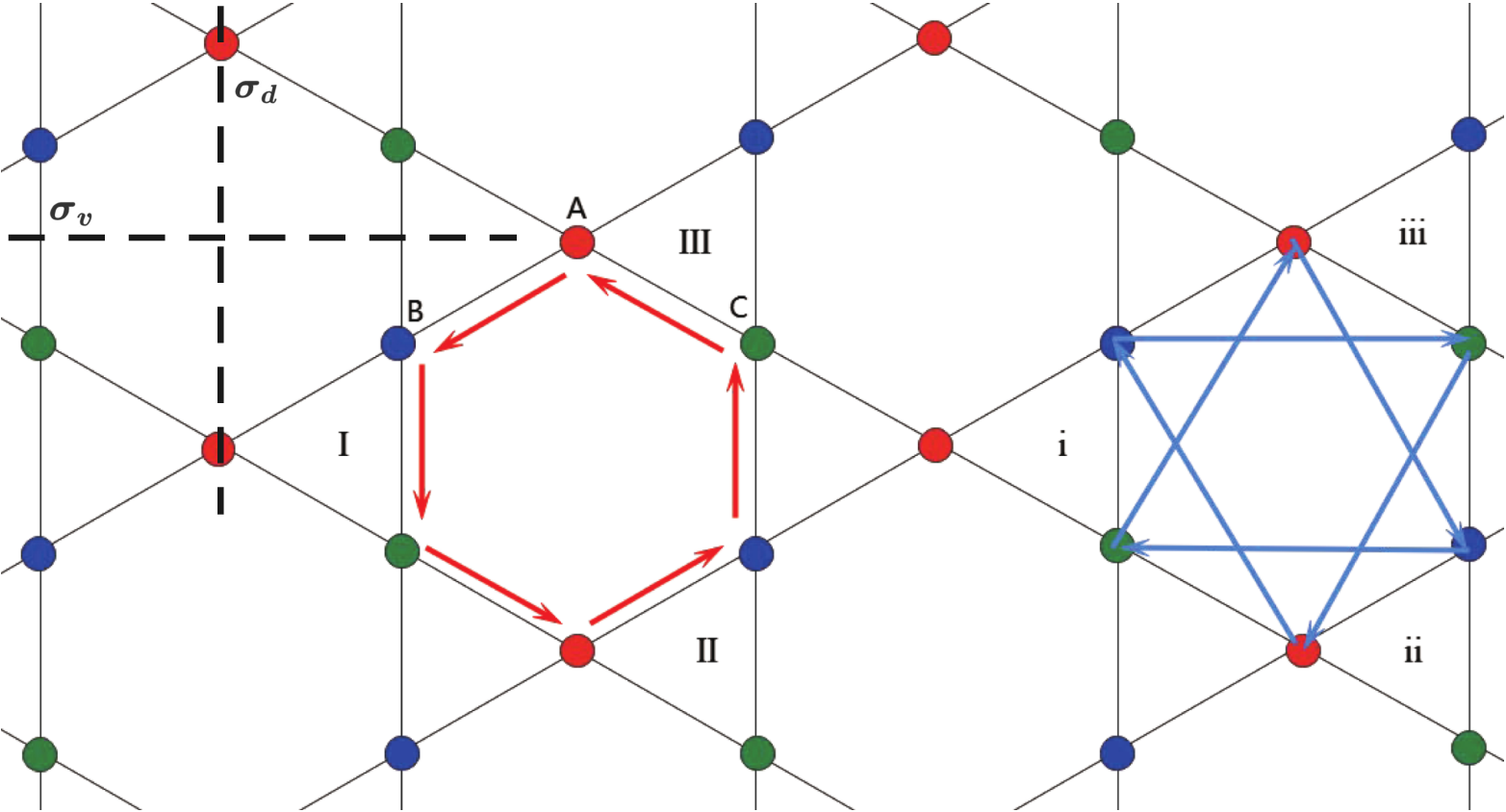}
	\caption{Schematic illustration of the kagome lattice. The red and blue vectors symbolize the NN and NNN bonds, respectively. The dashed lines indicate two non-equivalent mirror planes, $\sigma_d$ and $\sigma_v$, respectively.}
	\label{fig.1}
\end{figure}

The kagome lattice is illustrated in Fig.~\ref{fig.1}, with one unit cell containing three non-identical sites, depicted as red (A), blue (B) and green (C) points. As is seen, each site has four nearest-neighbor (NN) sites.
The normal state Hamiltonian includes the NN hopping term and the chemical potential term, expressed as
\begin{equation}
	H_N=-\sum_{\langle\textbf{i}\textbf{j}\rangle \sigma}(tc_{\textbf{i}\alpha\sigma}^{\dagger}c_{\textbf{j}\alpha^\prime\sigma}+H.c.)-\mu\sum_{\textbf{i} \sigma}c_{\textbf{i}\alpha\sigma}^{\dagger}c_{\textbf{i}\alpha\sigma},
	\label{eq2}
\end{equation}
with $t$ and $\mu$ being the NN hopping constant and the chemical potential, respectively.
$\alpha$ and $\sigma$ are the sublattice and spin indices, respectively. 
$\langle\textbf{ij}\rangle$ represents the NN bond. 

$H_{SC}$ represents the superconducting pairing term. Microscopically, superconducting pairing can be mediated by either electron-phonon interactions or electron-electron interactions.

In the case of phonon-mediated superconductors, the superconducting coherence length $L$ is typically quite large. Consequently, the pairing function generally consists of a superposition of all the $n$-th pairing states in real space, with $n$ increasing from $0$ to $L$. This usually leads to an isotropic pairing function in momentum space. Here, we primarily focus on the electronic structure near the Fermi energy. Assuming that the 
electronic structure is not sensitive to the
distortion of the gap function away from the Fermi energy, we approximate by considering a uniform $s$-wave pairing function in momentum space, with $\Delta({\bf k})\equiv \Delta_0$. This approach may provide a qualitative description of the phonon-mediated superconductor.

For electron-interaction mediated pairings, the superconducting pairing may occur over short ranges in real space and exhibit directionality within the lattice model. The superconducting pairing term in real space can be represented as follows:
\begin{equation}
	H_{SC}=\sum_{\textbf{i}\textbf{j}}(\Delta_{\textbf{ij}}c_{\textbf{i}\alpha\uparrow}^{\dagger}c_{\textbf{j}\alpha^\prime\downarrow}^{\dagger}+H.c.),
	\label{eq3}
\end{equation}
where ${\bf i}$ and ${\bf j}$ are the lattice sites of the two paired electrons. The phases of the superconducting order parameter might depend on the direction of the two pairing electrons, giving rise to distinct pairing symmetries. When transforming the pairing function into momentum space, an additional momentum-dependent factor is generally introduced.

Typically, for unconventional superconductors, the pairing states between nearest-neighbor (NN) sites or next-nearest-neighbor (NNN) sites are considered. Pairings involving longer distances between electrons result in higher-order pairing channels, which in turn lead to an increased number of nodal lines~\cite{PhysRevB.51.16233}. However, the presence of more nodal lines usually corresponds to higher free energy, making them less favorable. Consequently, longer-distance pairings are not considered.

We examine nine types of pairing symmetries in the context of superconductivity. These include the uniform $s$-wave pairing, as well as the pairing between NN sites and NNN sites. For both the NN-site pairing and the NNN-site pairing, four different pairing symmetries, namely, the $s$-wave pairing symmetry, the chiral $p+ip$-wave pairing symmetry, the chiral $d+id$-wave pairing symmetry, and the $f$-wave pairing symmetry, are examined.

For the NN-site pairing,
$\Delta_{\textbf{ij}}$ represents the superconducting pairing between the two NN sites, with $\Delta_{\textbf{ij}}=\Delta_{\textbf{i},\textbf{i}+\textbf{e}_j}=\Delta_j=\Delta_0e^{i\phi_j}$. The superconducting pairing symmetry is determined by the parity and the phase $\phi_j$. 
Given that a single unit cell encompasses three lattice sites, and each site possesses four NN sites, there is a total of twelve non-equivalent NN vectors. These comprise the six red vectors seen in Fig.~\ref{fig.1} and their respective inverse vectors. 
Then the order parameters of pairing, denoted as $\Delta_{\textbf{ij}}$, along these six vectors for different pairing symmetries are presented in Table \ref{tab:tabel1}.

\begin{table}[ht]
	\caption{\label{tab:tabel1}%
		The superconducting order parameters, corresponding to various red bond vectors within the NN pairing.}
	\begin{ruledtabular}
		\begin{tabular}{lcrrr}
			$\Delta_{\textbf{ij}}$ & $s$ & $p+ip$ & $d+id$ & $f$ \\
			
			\colrule
			$\Delta_{\rm{III_A,I_B}}$ & $\Delta_0$ & $\Delta_0$ & $\Delta_0$ & $\Delta_0$ \\
			
			$\Delta_{\rm{I_B,I_C}}$ & $\Delta_0$ & $e^{i\frac{\pi}{3}}\Delta_0$ & $e^{i\frac{2\pi}{3}}\Delta_0$ & $-\Delta_0$ \\
			
			$\Delta_{\rm{I_C,II_A}}$ & $\Delta_0$ & $e^{i\frac{2\pi}{3}}\Delta_0$ & $e^{i\frac{4\pi}{3}}\Delta_0$ & $\Delta_0$ \\
			
			$\Delta_{\rm{II_A,II_B}}$ & $\Delta_0$ & $-\Delta_0$ & $\Delta_0$ & $-\Delta_0$ \\
			
			$\Delta_{\rm{II_B,III_C}}$ & $\Delta_0$ & $e^{i\frac{4\pi}{3}}\Delta_0$ & $e^{i\frac{8\pi}{3}}\Delta_0$ & $\Delta_0$ \\
			
			$\Delta_{\rm{III_C,III_A}}$ & $\Delta_0$ & $e^{i\frac{5\pi}{3}}\Delta_0$ & $e^{i\frac{10\pi}{3}}\Delta_0$ & $-\Delta_0$ \\
		\end{tabular}
	\end{ruledtabular}
\end{table}

Similarly, for the NNN-site pairing term, twelve non-equivalent NNN bond vectors exist. The corresponding superconducting order parameters $\Delta_{\textbf{ij}}$ for various pairing symmetries along the six blue vectors are illustrated in Table \ref{tab:tabel2}. It's important to note that for both NN and NNN pairing, an additional constraint condition must be satisfied, with
$\Delta_{\textbf{ij}}=\xi\Delta_{\textbf{ji}}$, where $\xi$ is $+1$ for $s$-wave and $d+id$ pairing symmetries, and $\xi$ is $-1$ for $p+ip$ and $f$-wave pairing symmetries.

\begin{table}[ht]
	\caption{\label{tab:tabel2}The superconducting order parameters, corresponding to various blue bond vectors within the NNN pairing.}
	\begin{ruledtabular}
		\begin{tabular}{lcrrr}
			$\Delta_{\textbf{ij}}$ & $s$ & $p+ip$ & $d+id$ & $f$ \\
			\colrule
			$\Delta_{\rm{i_B,iii_C}}$ & $\Delta_0$ & $\Delta_0$ & $\Delta_0$ & $\Delta_0$ \\
			
			$\Delta_{\rm{i_C,iii_A}}$ & $\Delta_0$ & $e^{i\frac{\pi}{3}}\Delta_0$ & $e^{i\frac{2\pi}{3}}\Delta_0$ & $-\Delta_0$ \\
			
			$\Delta_{\rm{ii_A,i_B}}$ & $\Delta_0$ & $e^{i\frac{2\pi}{3}}\Delta_0$ & $e^{i\frac{4\pi}{3}}\Delta_0$ & $\Delta_0$ \\
			
			$\Delta_{\rm{iii_B,i_C}}$ & $\Delta_0$ & $-\Delta_0$ & $\Delta_0$ & $-\Delta_0$ \\
			
			$\Delta_{\rm{iii_C,ii_A}}$ & $\Delta_0$ & $e^{i\frac{4\pi}{3}}\Delta_0$ & $e^{i\frac{8\pi}{3}}\Delta_0$ & $\Delta_0$ \\
			
			$\Delta_{\rm{iii_A,ii_B}}$ & $\Delta_0$ & $e^{i\frac{5\pi}{3}}\Delta_0$ & $e^{i\frac{10\pi}{3}}\Delta_0$ & $-\Delta_0$ \\
		\end{tabular}
	\end{ruledtabular}
\end{table}

We take the distance between the NN sites as the length unit, then
we can transform the whole Hamiltonian into the momentum space, with,

\begin{equation}
	H=\sum_{\bf k}\Psi_{\textbf{k}}^{\dagger}\mathcal{H}_{\textbf{k}}\Psi_{\textbf{k}},
	\label{eq6}
\end{equation} 
where $\Psi_{\textbf{k}}^{\dagger}=(c_{\textbf{k}A\uparrow}^{\dagger},  c_{\textbf{k}B\uparrow}^{\dagger}, c_{\textbf{k}C\uparrow}^{\dagger},c_{-\textbf{k}A\downarrow},  c_{-\textbf{k}B\downarrow}, c_{-\textbf{k}C\downarrow})$, is the basis vector. 
$\mathcal{H}_\textbf{k}$ is a $6\times 6$ matrix, expressed as,
\begin{equation}
	\mathcal{H}_\textbf{k}=
	\begin{bmatrix}
		\mathcal{H}_0(\textbf{k})  & \Delta(\textbf{k})  \\
		\Delta^{\dagger}(\textbf{k}) & -\mathcal{H}_0(\textbf{k}).
	\end{bmatrix}
	\label{eq7}
\end{equation}
$\mathcal{H}_0(\textbf{k})$ is a $3\times 3$ matrix and is expressed as
\begin{equation}
	\mathcal{H}_0(\textbf{k})=
	\begin{pmatrix}
		-\mu    & -2t\cos k_1  & -2t\cos k_2 \\
		-2t\cos k_1  & -\mu    & -2t\cos k_3   \\
		-2t\cos k_2  & -2t\cos k_3 &-\mu   \\
	\end{pmatrix},
	\label{eq8}
\end{equation}
where $k_1=\frac{\sqrt{3}}{2}k_x+\frac{1}{2}k_y$, $k_2=\frac{\sqrt{3}}{2}k_x-\frac{1}{2}k_y$ and $k_3=-k_y$.

\begin{figure}
	\includegraphics[scale=0.5]{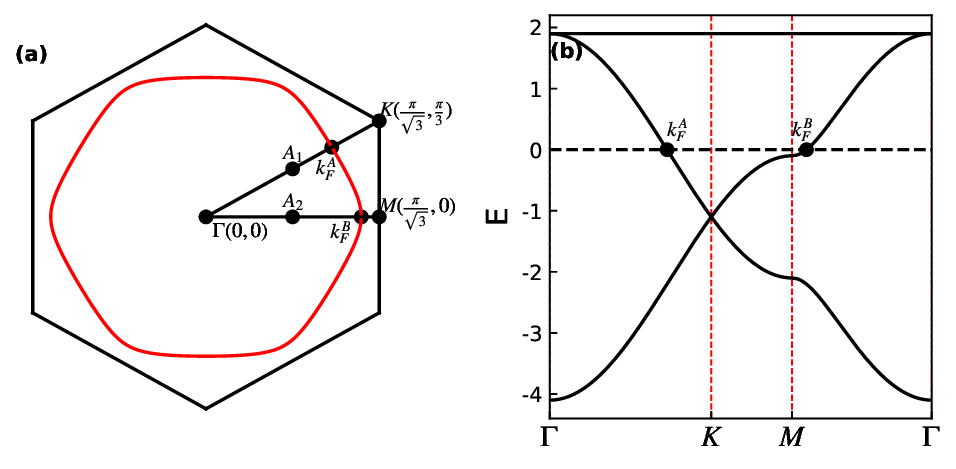}
	\caption{(a) The normal state Fermi surface. (b) The normal state energy bands along the highly symmetric lines.}
	\label{fig.2}
\end{figure}

The quasiparticle energy bands can be derived by diagonalizing the aforementioned matrix. With the chosen values of $t=1$ and $\mu=0.1$, the normal state Fermi surface, along with the energy bands along the highly symmetric lines, are depicted in Figs.~\ref{fig.2}(a) and \ref{fig.2}(b), respectively. The normal state Fermi surface constitutes a large pocket surrounding the $\Gamma$ point, with the upper Van Hove singularity positioned slightly below the Fermi energy. These characteristics are in line with the previously reported experimental findings and theoretical computations for kagome superconductors~\cite{r11,PhysRevB.106.014501,PhysRevB.105.174518,liu2023vanadiumbased}.

$\Delta(\textbf{k})$ is the $3\times 3$ superconducting pairing matrix. 
For the uniform $s$-wave pairing, $\Delta(\textbf{k})$ is expressed as,
\begin{equation}
	\Delta^{\mathrm{uni}}_{s}(\textbf{k})=
	\begin{pmatrix}
		\Delta_0 & 0 & 0  \\
		0 & \Delta_0 & 0  \\
		0 & 0 & \Delta_0  
	\end{pmatrix}.
	\label{eq9}
\end{equation}

For the NN pairing with four different pairing symmetries, the corresponding matrices $\Delta(\textbf{k})$ are obtained based on Table \ref{tab:tabel1},
 expressed as,
\begin{equation}
	\Delta^\mathrm{NN}_{s}(\textbf{k})=\Delta_0
	\begin{pmatrix}
		0 & 2\cos k_1 & 2\cos k_2  \\
		2\cos k_1 & 0 & 2\cos k_3  \\
		2\cos k_2 & 2\cos k_3 & 0
	\end{pmatrix},
	\label{eq10}
\end{equation}
\begin{equation}
	\begin{aligned}
		&\Delta^\mathrm{NN}_{p+ip}(\textbf{k})= \Delta_0 \\
		&\begin{pmatrix}
			0 & -2i\sin k_1 & -(\sqrt{3}+i)\sin k_2  \\
			-2i\sin k_1 & 0 & -(\sqrt{3}-i)\sin k_3  \\
			-(\sqrt{3}+i)\sin k_2 & -(\sqrt{3}-i)\sin k_3 & 0
		\end{pmatrix}
		\label{eq11}
	\end{aligned},
\end{equation}
\begin{equation}
	\begin{aligned}
		&\Delta^\mathrm{NN}_{d+id}(\textbf{k})=\Delta_0  \\
		&\begin{pmatrix}
			0 & 2\cos k_1 & -(1+\sqrt{3}i)\cos k_2  \\
			2\cos k_1 & 0 & -(1-\sqrt{3}i)\cos k_3  \\
			-(1+\sqrt{3}i)\cos k_2 & -(1-\sqrt{3}i)\cos k_3 & 0
		\end{pmatrix}
		\label{eq12}
	\end{aligned},
\end{equation}
and
\begin{equation}
	\Delta^\mathrm{NN}_{f}(\textbf{k})=\Delta_0
	\begin{pmatrix}
		0 & -2i\sin k_1 & 2i\sin k_2  \\
		-2i\sin k_1 & 0 & -2i\sin k_3  \\
		2i\sin k_2 & -2i\sin k_3 & 0
	\end{pmatrix}.
	\label{eq13}
\end{equation} 

For the NNN pairing, according to Table \ref{tab:tabel2},
the corresponding $\Delta(\textbf{k})$ matrices are expressed as,
\begin{equation}
	\Delta^\mathrm{NNN}_{s}(\textbf{k})=\Delta_0
	\begin{pmatrix}
		0 & 2\cos k_4 & 2\cos k_5  \\
		2\cos k_4 & 0 & 2\cos k_6  \\
		2\cos k_5 & 2\cos k_6 & 0
	\end{pmatrix},
	\label{eq14}
\end{equation}
\begin{equation}
	\begin{aligned}
		&\Delta^\mathrm{NNN}_{p+ip}(\textbf{k})=\Delta_0  \\
		&\begin{pmatrix}
			0 & (\sqrt{3}+i)\sin k_4 & (\sqrt{3}-i)\sin k_5  \\
			(\sqrt{3}+i)\sin k_4 & 0 & 2i\sin k_6  \\
			(\sqrt{3}-i)\sin k_5 & 2i\sin k_6 & 0
		\end{pmatrix},
		\label{eq15}
	\end{aligned}
\end{equation}
\begin{equation}
	\begin{aligned}
		&\Delta^\mathrm{NNN}_{d+id}(\textbf{k})=\Delta_0  \\
		&\begin{pmatrix}
			0 & -(1+\sqrt{3}i)\cos k_4 & -(1-\sqrt{3}i)\cos k_5  \\
			-(1+\sqrt{3}i)\cos k_4 & 0 & 2\cos k_6  \\
			-(1-\sqrt{3}i)\cos k_5 & 2\cos k_6 & 0
		\end{pmatrix}
		\label{eq16}
	\end{aligned},
\end{equation}
and
\begin{equation}
	\Delta^\mathrm{NNN}_{f}(\textbf{k})=\Delta_0
	\begin{pmatrix}
		0 & -2i\sin k_4 & 2i\sin k_5  \\
		-2i\sin k_4 & 0 & 2i\sin k_6  \\
		2i\sin k_5 & 2i\sin k_6 & 0
	\end{pmatrix},
	\label{eq17}
\end{equation}
with $k_4=\frac{\sqrt{3}}{2}k_x-\frac{3}{2}k_y$, $k_5=-\frac{\sqrt{3}}{2}k_x-\frac{3}{2}k_y$ and $k_6=\sqrt{3}k_x$. \par

The bare Green's function matrix [$\hat{G}_0(\textbf{k},\omega)$] in the momentum space can be obtained through diagonalizing the above Hamiltonian. It is defined as $\hat{G}_0(\textbf{k},\omega)$, with the elements being expressed as, 
\begin{equation}
	G_{0ij}(\textbf{k},\omega)=\sum_{n}\frac{u_{i,n}(\textbf{k})u_{j,n}^{*}(\textbf{k})}{\omega-E_n(\textbf{k})+i\Gamma},
	\label{eq19}
\end{equation}
where $u_{in}(\textbf{k})$ and $E_n({\bf k})$ are the eigenvectors and the eigenvalue of the Hamiltonian matrix, respectively. 

Quasiparticle excitations can, in principle, be described by the imaginary parts of the Green's function. In particular, the spectral function can be obtained from the diagonal elements of the Green's function and is expressed as: 
\begin{equation}
	A(\textbf{k},\omega)=-\frac{1}{\pi} {\rm Im} \sum^3_{i=1}[G_{0ii}(\textbf{k},\omega)+[G_{0i+3,i+3}(\textbf{k},-\omega)].
	\label{eq23}
\end{equation}

The information about the superconducting pairing can be obtained from the off-diagonal elements of the Green's function (also called the anomalous Green's function). We define a momentum-dependent pairing function $F({\bf k})$ by integrating all the occupied quasiparticle states as follows:
\begin{equation}
F({\bf k})=\int_{-\infty}^0\mathrm{Im}\sum^3_{i=1}G_{0i,i+3}({\bf k},\omega)d\omega.
\end{equation}

We consider an additional point impurity to study the impurity effect. The additional Hamiltonian contributed by the impurity is expressed as,
\begin{equation}
	H_{imp}=V_s\sum_{\sigma}c_{\textbf{r}_0\alpha\sigma}^{\dagger}c_{\textbf{r}_0\alpha\sigma},
	\label{eq18}
\end{equation}
with $V_s$ being the impurity strength. $\textbf{r}_0$ is the site where the impurity locates.

Then the T-matrix and the full Green's function are expressed as~\cite{RevModPhys.78.373},
\begin{equation}
	\hat{T}(\omega)=\frac{\hat{U}}{\hat{I}-\hat{U}\hat{G}_0(\omega)}
	\label{eq20}
\end{equation}
and
\begin{equation}
	\hat{G}(\textbf{r},\textbf{r}',\omega)=\hat{G}_0(\textbf{r},\textbf{r}',\omega)+\hat{G}_0(\textbf{r},0,\omega)\hat{T}(\omega)\hat{G}_0(0,\textbf{r}',\omega).
	\label{eq21}
\end{equation}
Here $\hat{G}_0(\textbf{r},\textbf{r}',\omega)=\frac{1}{N}\sum\limits_{k}\hat{G}_0(\textbf{k},\omega)e^{i\textbf{k}\cdot(\textbf{r}-\textbf{r}')}$ is the Fourier transformation of $\hat{G}_0(\textbf{k},\omega)$. $\hat{I}$ is the identity matrix. We put a point impurity on the sublattice A. $\hat{U}$ is a $6\times 6$ matrix with $U_{11}=-U_{44}=V_s$.
The LDOS near the impurity can be obtained as 
\begin{equation}
	\rho(\textbf{r},\omega)=-\frac{1}{\pi}{\rm Im} [G_{22}(\textbf{r},\textbf{r},\omega)+G_{55}(\textbf{r},\textbf{r},-\omega)].
	\label{eq22}
\end{equation}

\section{\label{sec3.rad}RESULTS AND DISCUSSION}

\subsection{Group classification of superconducting states}

The superconducting pairing symmetries can be classified based on the general symmetry classification~\cite{PhysRevB.51.16233,RevModPhys.72.969,PhysRevB.90.054521,PhysRevB.92.085121,PhysRevB.100.214507}. The superconducting phase transition from the normal state to the superconducting state is associated with symmetry breaking, which can be described using the superconducting order parameter. This order parameter can be classified according to the basis functions of the irreducible representations for the symmetry group of the crystal lattice.

In the case of a two-dimensional kagome lattice, it belongs to the $C_{6v}$ point group which has twelve elements in six conjugacy classes. This includes a six-fold rational symmetry and two non-equivalent mirror planes, $\sigma_d$ and $\sigma_v$, as shown in Fig.~\ref{fig.1}. There are six non-equivalent irreducible representations for the $C_{6v}$ point group, and their character table is presented in Table III.

\begin{table}[ht]
	\caption{\label{tab:tabel3}The character table of the $C_{6v}$ point group.}
	\begin{ruledtabular}
		\begin{tabular}{lcccccc}
			$C_{6v}$ & $E$ & $2C^{\pm}_6$ & $2C^{\pm}_3$ & $C_2$ & $3\sigma_v$ &   $3\sigma_d$\\
			\colrule
			$A_1$ & 1 & 1 & 1 & 1 &1 &1\\
			
			$A_2$ & 1 & 1 & 1 & 1 &-1 &-1 \\
			
			$B_1$ & 1 & -1 & 1 & -1 &1 &-1  \\
			
			$B_2$ & 1 & -1 & 1 & -1 &-1 & 1  \\
			$E_1$ & 2 & 1 & -1 & -2 &0 &0  \\
			$E_2$ & 2 & -1 & -1 & 2 &0 &0  \\
			
		\end{tabular}
	\end{ruledtabular}
\end{table}

$E_1$ and $E_2$ are the two-dimensional irreducible representations for this point group. Each element in these representations is represented by a two-dimensional matrix, which can be found in Table IV.

\begin{table*}
	\caption{\label{tab:tabel4}The $E_1$ and $E_2$ irreducible representations.}
	\begin{ruledtabular}
		\begin{tabular}{lcccccc}
			$C_{6v}$ & $E$ & $2C^{\pm}_6$ & $2C^{\pm}_3$ & $C_2$ & $3\sigma_v$ &   $3\sigma_d$\\ 
			\colrule

			\vspace{2pt}
			$E_1$ & $\left[ {\begin{array}{cc} 1 & 0\\ 0&1 \end{array}} \right]$ & $\left[ {\begin{array}{cc} \frac{1}{2} & \mp \frac{\sqrt{3}}{2}\\ \pm \frac{\sqrt{3}}{2}& \frac{1}{2} \end{array}} \right]$ &$\left[ {\begin{array}{cc} -\frac{1}{2} & \mp \frac{\sqrt{3}}{2}\\ \pm \frac{\sqrt{3}}{2}& -\frac{1}{2} \end{array}} \right]$ &$\left[ {\begin{array}{cc} -1 & 0\\ 0&-1 \end{array}} \right]$ & $\left[ {\begin{array}{cc} -1 & 0\\ 0&1 \end{array}} \right]$ &$\left[ {\begin{array}{cc} 1 & 0\\ 0&-1 \end{array}} \right]$ \\

\vspace{2pt}
		
$E_2$ & $\left[ {\begin{array}{cc} 1 & 0\\ 0&1 \end{array}} \right]$ & $\left[ {\begin{array}{cc} -\frac{1}{2} & \pm \frac{\sqrt{3}}{2}\\ \mp \frac{\sqrt{3}}{2}& -\frac{1}{2} \end{array}} \right]$ &$\left[ {\begin{array}{cc} -\frac{1}{2} & \mp \frac{\sqrt{3}}{2}\\ \pm \frac{\sqrt{3}}{2}& -\frac{1}{2} \end{array}} \right]$ &$\left[ {\begin{array}{cc} 1 & 0\\ 0&1 \end{array}} \right]$ & $\left[ {\begin{array}{cc} 1 & 0\\ 0&-1 \end{array}} \right]$ &$\left[ {\begin{array}{cc} 1 & 0\\ 0&-1 \end{array}} \right]$

		\end{tabular}
	\end{ruledtabular}
\end{table*}

The simplest basis functions of the irreducible representations can be obtained through Tables III and IV, and they are listed in the second column of Table V. The corresponding order parameter phase illustrations in the first Brillouin zone based on these simplest basis functions are given in the third column. Among them, the simplest basis function of the $A_{2}$ irreducible representation has multiple nodal lines, leading to a higher free energy for the $A_{2}$ superconducting state. This state can only be obtained with long-range pairing interactions and is unlikely to be realized. Therefore, it is not discussed in this work. For the $E_1$ and $E_2$ irreducible representations, the basis functions have two components, and the actual pairing order parameter should be a linear combination of them. Generally, chiral pairing states with a $\pi/2$ phase shift between these two components are favorable~\cite{PhysRevB.90.054521}.

\begin{table}
\caption{The simplest basis functions of the irreducible representations and symmetry schematic in the first Brillouin zone from the basis functions.}
\begin{tabular}{ccc}
\hline \hline
 $C_{6v}$ & Simplest basis functions & Symmetry schematic\\
  
 \hline
~\\
 \vspace{4pt}
 
 $A_{1}$ & 1 &
 \parbox[c]{0em}{\includegraphics[height =1.4cm]{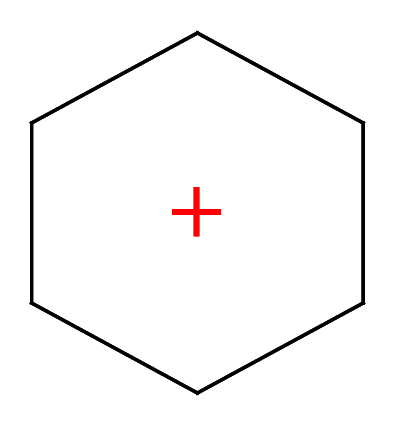} }\\
\vspace{4pt}

 $A_{2}$ & $k_x k_y (k_x^2 - 3k_y^2)(k_y^2 - 3k_x^2)$ & 
 
  \parbox[c]{0em}{\includegraphics[height =1.4cm]{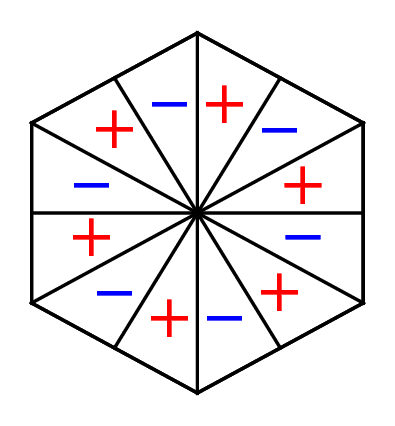} } \\

\vspace{4pt}

$B_{1}$ & $k_x(k_x^2-3k_y^2)$ &
 \parbox[c]{0em}{\includegraphics[height =1.4cm]{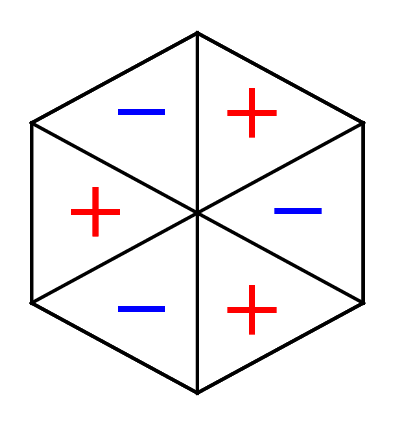} } \\
\vspace{4pt}
$B_{2}$ & $k_y(k_y^2-3k_x^2)$ &
 \parbox[c]{0em}{\includegraphics[height =1.4cm]{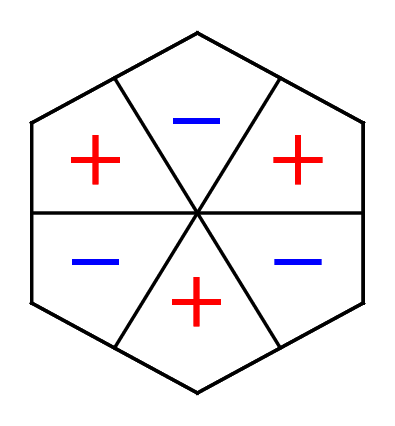} } \\
\vspace{4pt}
$E_{1}$ & $(k_x, k_y)$ &
 \parbox[c]{0em}{\includegraphics[height =0.9cm]{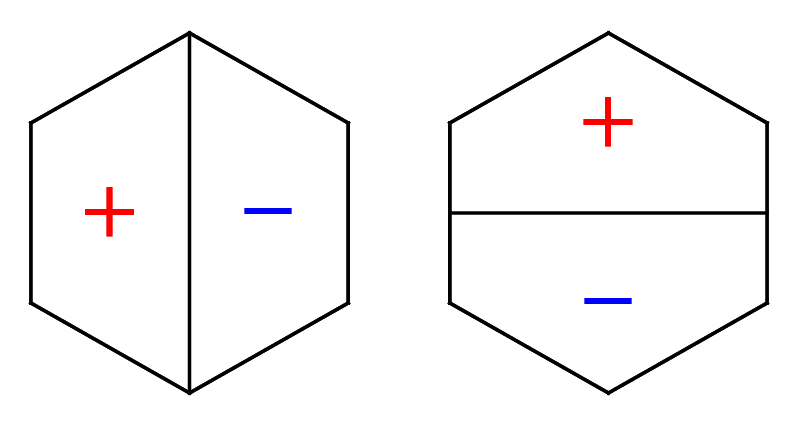} } \\
\vspace{4pt}
   $E_{2}$ & $(2k_x k_y,k_x^2-k_y^2)$ &
 \parbox[c]{0em}{\includegraphics[height =0.9cm]{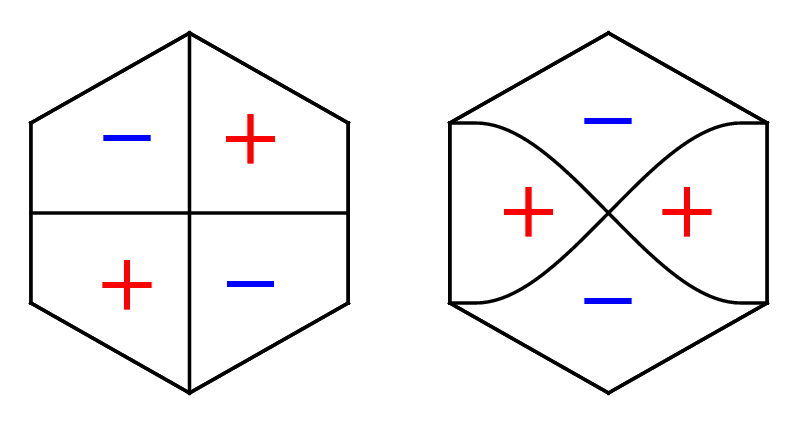} } \\

  \hline \hline
 \end{tabular}
  
 \label{tab:sym}
\end{table}

\begin{figure} 
	\centering 
	\includegraphics[scale=0.6]{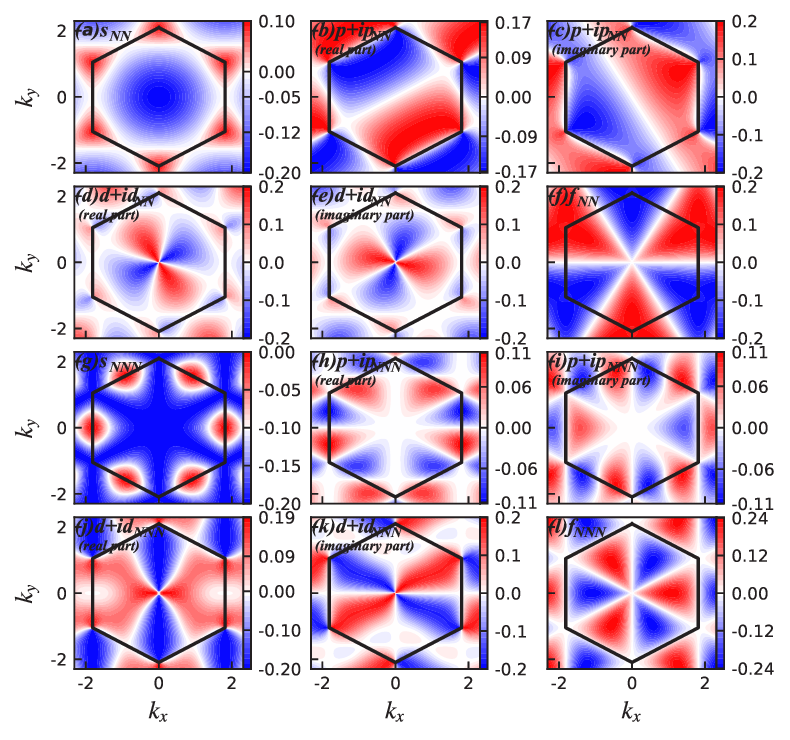} 
	\caption{The intraband pairing functions for the eight unconventional pairing channels. For the $p+ip$ and $d+id$ pairing symmetries, the pairing functions are complex including the real parts and the imaginary parts, which are presented separately.} 
\label{fig.3}
\end{figure}

We here obtain the pairing functions by defining the pairing terms in the real space of the kagome lattice. This method ensures that the obtained pairing functions in the momentum space are naturally associated with the basis functions of the irreducible representations~\cite{PhysRevB.92.085121,PhysRevB.100.214507}. The uniform $s$-wave pairing symmetry generates a uniform superconducting pairing and does not break any lattice symmetry, thus belonging to the $A_{1}$ irreducible representation. For the other eight unconventional pairing channels, possible anisotropic energy gaps may exist. The anisotropic behaviors can be observed more clearly by exploring the pairing functions in the momentum space.

In the momentum space, there are three normal state energy bands in the normal state, as seen in Fig.~\ref{fig.2}. The pairing functions are expressed as the $3\times 3$ matrices [Eqs.(7-15)]. To demonstrate the symmetries of the pairing functions, the pairing function matrices are transformed into the band basis with $\tilde{\Delta_{\bf k}}=V_{\bf k}^\dagger \Delta_{\bf k} V_{\bf k}$, where $V_{\bf k}$ is the eigenvector matrix of the $3\times 3$ normal state matrix $\mathcal{H}_0(\textbf{k})$. The diagonal components of the matrix $\mathcal{H}_0(\textbf{k})$ reflect the intra-band pairing. For the kagome superconductor, band $2$ is important due to the existence of the Van Hove singularity points near the Fermi energy in this band, as shown in Fig.~\ref{fig.2}.

The numerical results of intraband pairing functions for the eight unconventional pairing channels in the momentum space for band $2$ are presented in Fig.~\ref{fig.3}. The symmetries of the different pairing channels are clearly reflected, and the obtained pairing symmetries are indeed associated with irreducible representations. They are consistent with the basis functions of irreducible representations. The $s$-wave pairing symmetries belong to the $A_{1}$ irreducible representation. The $f_{NN}$ pairing symmetry and the $f_{NNN}$ one belong to the $B_2$ and $B_1$ irreducible representations, respectively. The $p+ip$ pairing symmetries and $d+id$ pairing symmetries belong to the $E_1$ and $E_2$ irreducible representations, respectively.

For the NNN $p+ip$ and $d+id$ pairing symmetries, the overall symmetric behavior is consistent with the basis functions of the $E_1$ and $E_2$ irreducible representations. However, additional nodal lines are induced, indicating the existence of a higher-order pairing term. In principle, the higher-order pairing term may be induced by the long-range pairing interaction and then more nodal lines may exist~\cite{PhysRevB.51.16233}. Numerically, it has been checked that when farther pairing terms are considered, more nodal lines will be induced.

\begin{figure} 
	\centering 
	\includegraphics[scale=0.6]{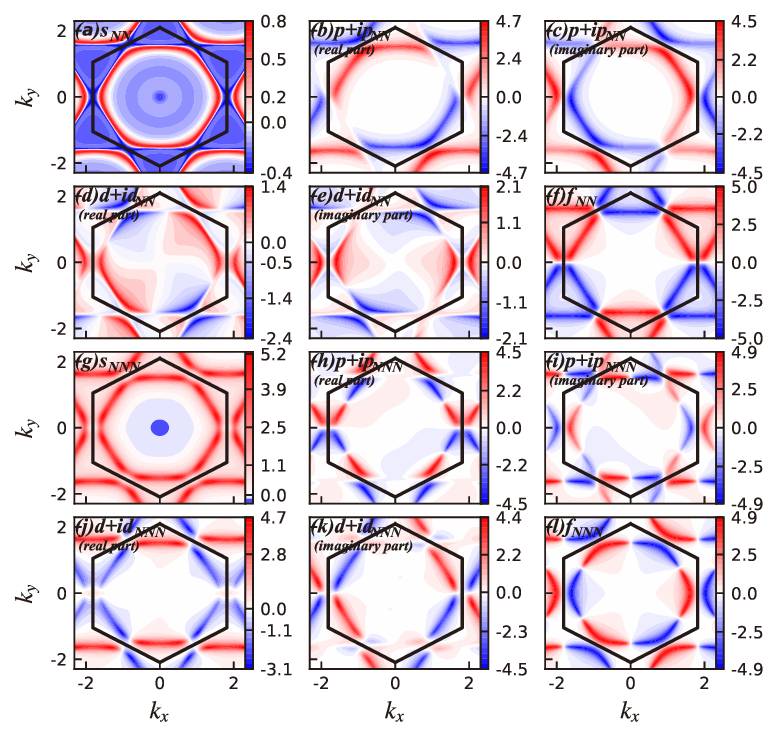} 
	\caption{Similar to Fig.3 but the pairing functions are obtained from the anomalous Green's function [Eq.(18)].}  
\label{fig.4}
\end{figure}

Exploring the pairing symmetry based on the momentum-dependent anomalous Green's function is also useful. The intensity plots of the pairing function obtained by the anomalous Green's function [defined by Eq.(18)] in the momentum space for all of the eight unconventional pairing channels are presented in Fig.~\ref{fig.4}. The symmetric features of the anomalous Green's functions are generally consistent with the pairing functions shown in Fig.~\ref{fig.3}. However, the anomalous Green's functions are renormalized by the quasiparticle energy bands, causing them to decrease significantly when moving away from the normal state Fermi surface. They can reflect the pairing order parameters around the Fermi surface.

\subsection{Single particle spectra}

\begin{figure} 
	\centering 
	\includegraphics[scale=0.5]{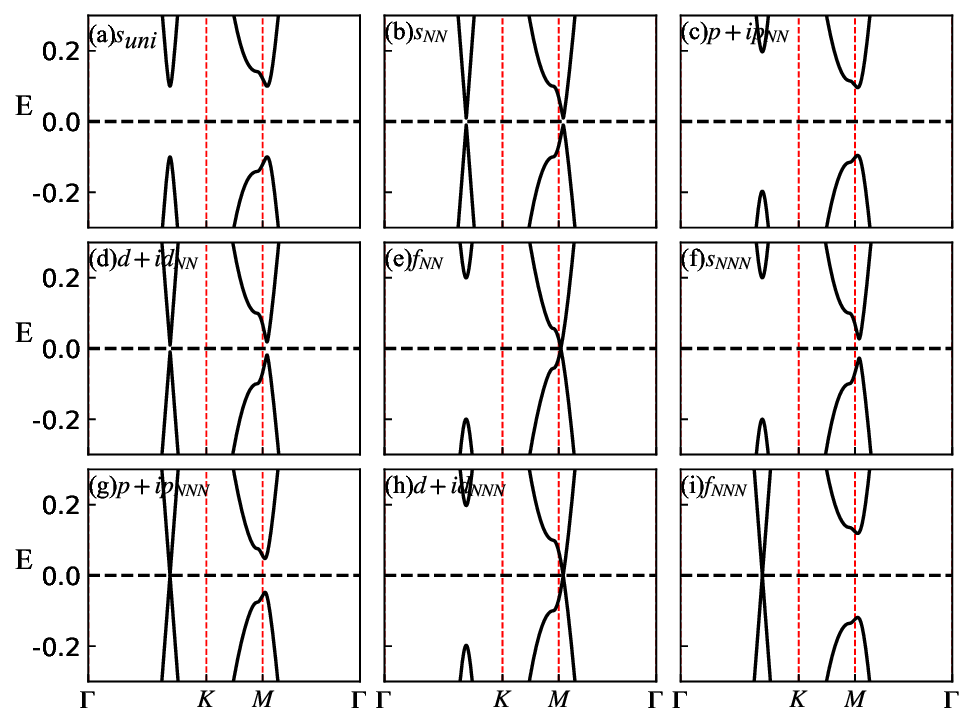} 
	\caption{Superconducting energy bands along highly symmetric lines for the uniform $s$-wave pairing state ($s_{uni}$) and eight unconventional pairing states.} 
	\label{fig.5} 
\end{figure}

The pairing symmetry for the superconductor cannot be directly obtained through experiments. However, useful information can be acquired by studying the electronic structure in the superconducting state.
In this context, let's consider the single particle spectra in various superconducting states that exhibit different pairing symmetries. After the diagonalization of the $6\times 6$ superconducting Hamiltonian [Eq.(\ref{eq7})] with $\Delta_0=0.1$, the energy bands along the symmetric lines are displayed in Fig.~\ref{fig.5}. It is important to note that, in the normal state, the energy bands along the $\Gamma-K$ and $M-\Gamma$ lines intersect the Fermi energy at the Fermi momentum coordinates $k^A_F$ and $k^B_F$, respectively [Fig.~\ref{fig.2}].
In the superconducting state, an energy gap may appear at moments $k^A_F$ or $k^B_F$. As shown in Fig.~\ref{fig.5}, the energy bands of the  $s$-wave pairing symmetry, the NN $p+ip$ pairing symmetry, and the NNN $s$-wave pairing symmetry are fully gapped. The energy gap for the uniform $s$-wave pairing symmetry is isotropic, with $\Delta(k^A_F)=\Delta(k^B_F)=\Delta_0$. However, for the NN $p+ip$ pairing symmetry and the NNN $s$-wave pairing symmetry, the energy gap is anisotropic, where $\Delta(k^A_F)>\Delta(k^B_F)$.

The numerical findings also reveal that, for the NN $s$-wave pairing symmetry and NN $d+id$ pairing symmetry, the system is fully gapped with a slight energy gap. In the NN $s$-wave pairing symmetry, a gap of about $0.1\Delta_0$ occurs at both Fermi momentum points. For the NN $d+id$ pairing symmetry, the energy gap is approximately $0.1\Delta_0$ at the momentum $k^A_F$, and about $0.18\Delta_0$ at the momentum $k^B_F$.

The energy bands for the NN $f$-wave, NNN $p+ip$ pairing symmetry, NNN $d+id$ pairing symmetry, and NNN $f$-wave pairing symmetry display nodal points. For the NN $f$-wave pairing symmetry and the NNN $d+id$ pairing symmetry, a sizable energy gap exists at the momentum $k^A_F$, while the gap closes at the momentum $k^B_F$. The energy gaps of the NNN $p+ip$ pairing symmetry and the NNN $f$-wave pairing symmetry manifest at the momentum $k^B_F$, and disappear at the momentum $k^A_F$. These quantitative results strongly suggest that the quasiparticle energy bands are heavily dependent on the pairing symmetries and can be used to distinguish different pairing symmetries.

\begin{figure} 
	\centering 
	\includegraphics[scale=0.55]{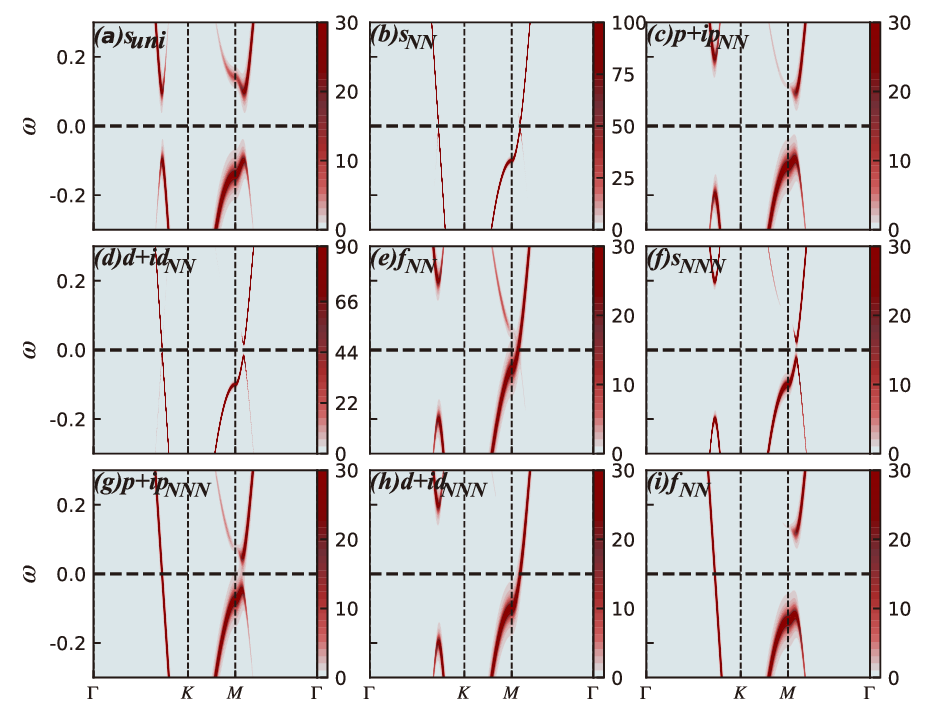} 
	\caption{The intensity plots of the spectral function as functions of the energy and the momentum along the highly symmetric lines for different superconducting pairing states.} 
	\label{fig.6} 
\end{figure}

The numerical results of energy bands qualitatively agree with the pairing functions depicted in Figs.~\ref{fig.3} and ~\ref{fig.4}. In experimental terms, quasiparticle energy bands can be directly explored using ARPES experiments. On theoretical grounds, the ARPES spectra can be defined utilizing the spectral function. The intensity plots of the spectral functions, in relation to the momentum and the energy, along the highly symmetric lines are shown in Fig.~\ref{fig.6}. The quasiparticle dispersions for various pairing symmetries are vividly depicted. It is noteworthy that, the energy bands in the superconducting state have the particle-hole symmetry, however, the spectral weight for each band is distinct and hence, only fragments of energy bands can be clearly discerned in the spectral function spectra. The potential energy gaps at the Fermi level for typical pairing symmetries can also be identified through the spectral functions. Substantial consistency is observed in the corresponding results between the energy bands and the spectral functions shown in Figs.~\ref{fig.5} and ~\ref{fig.6}.

\begin{figure} 
	\centering 
	\includegraphics[scale=0.6]{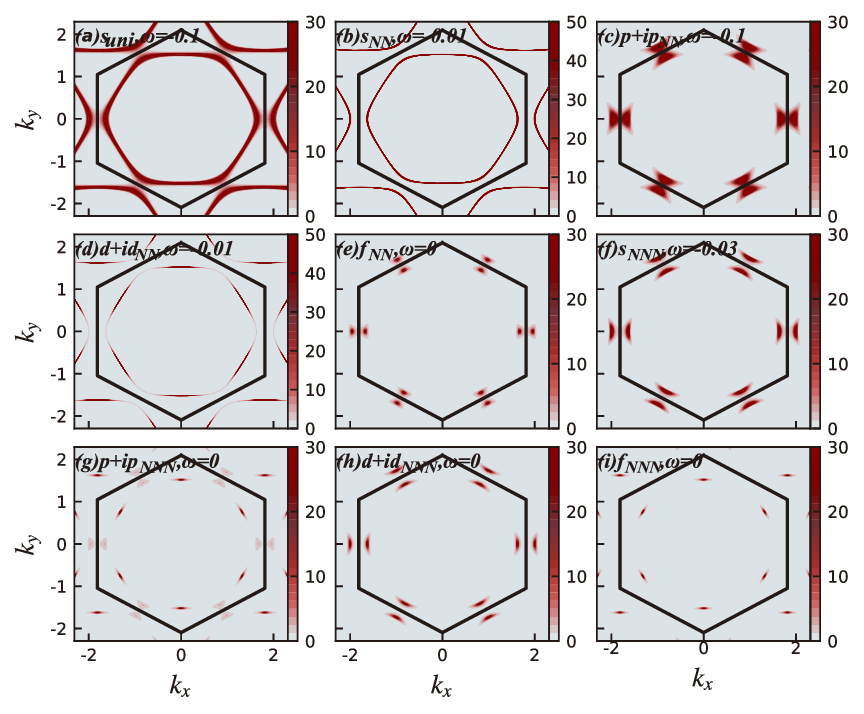} 
	\caption{The intensity plots of the spectral function as a functions of the momentum at the constant energy for different superconducting pairing states. The black hexagons are the first Brillouin zone.} 
	\label{fig.7} 
\end{figure}

It is also informative to evaluate the spectral functions at a constant energy to investigate the momentum dependence of the quasiparticle excitation in the whole Brillouin zone. We are interested in the low energy quasiparticle excitation. When the energy bands display nodal points, we exhibit the zero energy spectral functions. For fully gapped superconducting states, energies are chosen as the minimum energy of the quasiparticle. Subsequently, we illustrate the intensity plots of the spectral functions as a function of momentum at a constant energy in Fig.~\ref{fig.7}. Three qualitatively different spectra are revealed. For the uniform $s$-wave and the NN $s$-wave pairing symmetries. The low energy quasiparticle excitation emerges along the entire Fermi surface, indicating that the superconducting gap is isotropic along the entire Fermi surface. For the NN $p+ip$ pairing, the NN $f$-wave pairing, the NNN $s$-wave pairing, and the NNN $d+id$ pairing, the quasiparticle excitations emerge near the $M$ point (at the momentum $k^B_F$ and its symmetric points), designating the superconducting gap at momentum $k^B_F$ to be miniscule or nonexistent. For the other three pairing symmetries (the NN $d+id$ pairing, the NNN $p+ip$ pairing, and the NNN $f$-wave pairing), the quasiparticle excitations emerge around the momentum $k^A_F$ and its symmetric points, implying that for these pairing symmetries, the superconducting gap at the momentum $k^A_F$ is miniscule or zero. These findings are qualitatively consistent with the numerical results along the highly symmetric lines presented in Figs.~\ref{fig.5} and \ref{fig.6}.

\begin{figure} 
	\centering 
	\includegraphics[scale=0.5]{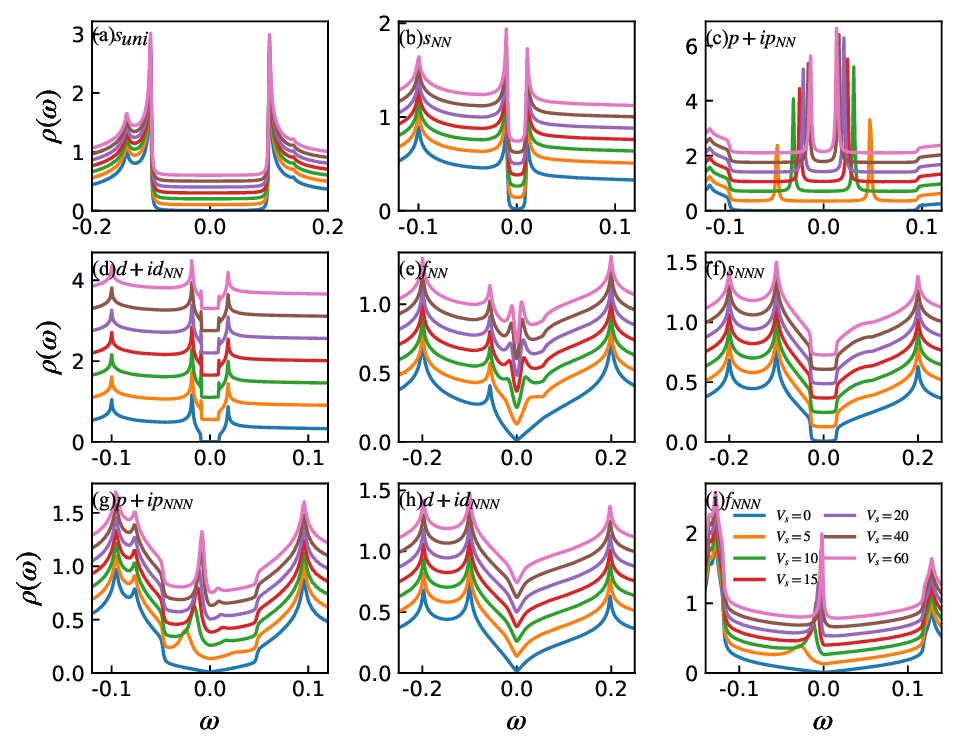} 
	\caption{The LDOS spectra at the NN site of the impurity with different impurity strengths $V_s$ and different pairing symmetries. From the bottom to the top, $V_s$ increases from 0 to 60.} 
	\label{fig.8} 
\end{figure}

We now focus on studying the numerical results of the LDOS. The LDOS spectra at the NN site to the impurity for different pairing states with varying impurity strengths ($V_s$) are shown in Fig~.\ref{fig.8}. Without the impurity ($V_s=0$), the LDOS spectra can be used to ascertain whether the system is fully gapped. If the system is indeed fully gapped, the LDOS at low energies equates to zero and the spectrum manifests in a 'U' shape. If nodal points exist, the LDOS will not completely disappear at any non-zero energy and the spectrum adopts a 'V' shape. Our numerical results suggest that for the NN $f$-wave state, NNN $p+ip$ state, NNN $d+id$ state, and NNN $f$-wave state, the LDOS spectrum creates a 'V' shape, implying that with these pairing symmetries, the system is not fully gapped, congruent with the numerical results of the quasiparticle energy bands.

Let us examine the influence of impurity effects. As shown in Fig~.\ref{fig.8}, the effects of impurity indeed heavily depend on the pairing symmetries. For all the three $s$-wave states discerned, no in-gap states exist. Two subdued in-gap peaks near the gap edge are induced by the impurity for the NN $d+id$ state. For the NNN $d+id$ pairing state, some additional in-gap states are brought forth by the impurity, specifically, a sizeable bump exists at a positive energy. For the two spin triplet pairing states ($p+ip$ state or $f$-wave state), there are sharp resonant peaks inside the energy gap. The existing sharp resonant peaks might serve as a distinctive characteristic for the spin triplet pairing in the kagome superconductor.

Theoretically, the impurity effects are responsive to the phase of the superconducting order parameter. The resonant states generally emerge when the sign of the superconducting gap changes based on the Andreev reflection picture~\cite{PhysRevLett.72.1526,PhysRevLett.103.186402,PhysRevB.80.064513}. Here, as seen in Tables \ref{tab:tabel1} and \ref{tab:tabel2}, for the three $s$-wave states, no sign changes occur, whereas, for the $p+ip$ pairing and $f$-wave pairing, the sign change generally arises due to the properties of the spin triplet pairing thereby creating sharp in-gap resonant states. For the the $d+id$ pairing, the phase of the superconducting gap changes, generating the in-gap states, while in this case there are no sign changes occurring thus no strong resonant states exist.

\begin{figure}
	\centering
	\includegraphics[scale=0.5]{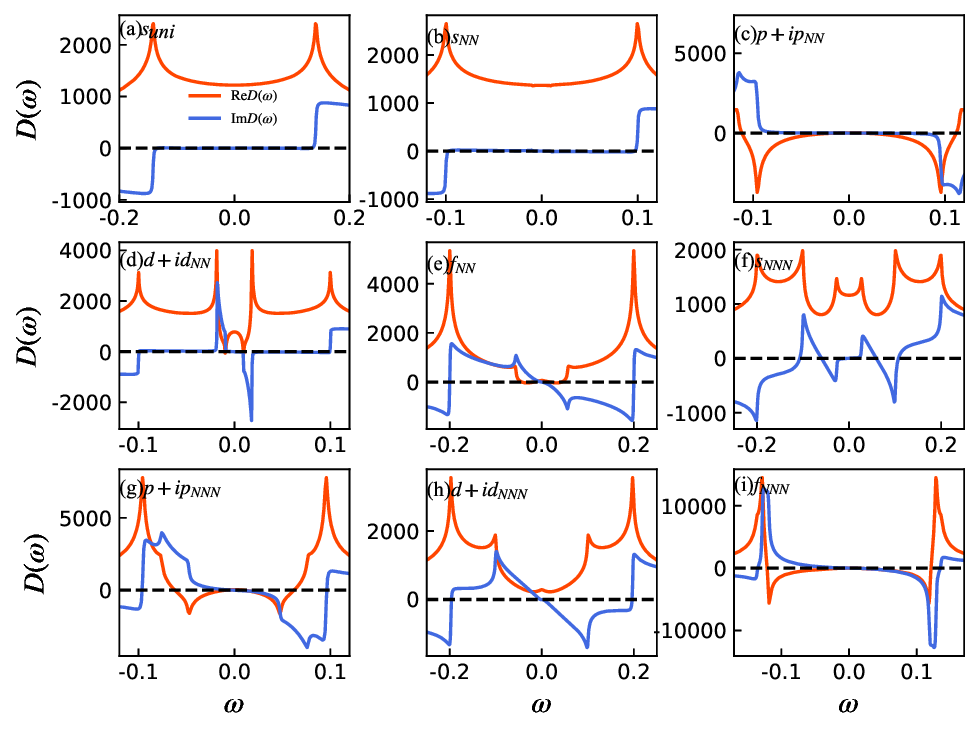}
	\caption{The real and imaginary parts of $D(\omega)$ as a function of the energy $\omega$ with the impurity strength $V_s=60$.}
	\label{fig.9}
\end{figure}

An alternative method to explain the existence of these in-gap states can be derived from an analysis of the denominator of the T-matrix $D(\omega)$, where $D(\omega) = |\hat{I}-\hat{U}\hat{G}_0(\omega)|$. As the energy resides inside the superconducting gap, the imaginary part of $D(\omega)$ is generally small. A resonant state emerges when its real part approaches zero at a particular low energy. The real and imaginary parts of $D(\omega)$ with $V_s=60$ for different pairing symmetries are shown in Fig.~\ref{fig.9}. The fundamental features of the impurity effect illustrated in Fig.~\ref{fig.8} can be fully accounted for. For both the $p+ip$ pairing states and the $f$-wave pairing states, the real and imaginary parts of $D(\omega)$ tend to zero near the Fermi energy, resulting in resonant states. For the NN $d+id$ pairing state, the real part of $D(\omega)$ intersects the zero axis at an energy near the gap edge, whereas at the gap edge, the absolute value of the imaginary part, $\mid$ Im$D(\omega)$ $\mid$, generally increases in a steplike manner due to the existence of the superconducting coherent peak. As a result, the impurity-induced in-gap peak is quite weak. For the NN $d+id$ pairing state, the real part of $D(\omega)$ does not cross the zero axis, while it is minimal at low energies, leading to the in-gap bump structure. For the three $s$-wave pairing states, the real part of $D(\omega)$ is consistently large, thereby resulting in no in-gap states for these pairing states.

Our numerical results indicate that single-particle spectra are heavily influenced by pairing symmetries, making them effective tools for distinguishing between different pairing states. Comparing our theoretical findings with recent experiments on kagome superconductors provides valuable insights.

The AV$_3$Sb$_5$ family has attracted considerable experimental interest. Impurity effects in the AV$_3$Sb$_5$ family have been investigated using STM experiments. In these experiments, no in-gap states were detected in the superconducting state~\cite{r23}, initially ruling out the possibilities of $p+ip$ and $f$-wave pairing for this family. A recent ARPES experiment~\cite{Zhong2023} reported an isotropic superconducting gap, with the energy gap around the Fermi surface ($\Delta$) and the superconducting critical temperature $T_c$ yielding a ratio of $2\Delta/K_B T_c \approx 3.44$. This ratio closely aligns with that of a conventional superconductor, suggesting that the AV$_3$Sb$_5$ family may be phonon-mediated superconductors. As previously mentioned, the electronic structure is assumed to be insensitive to the distortion
of the gap function and the uniform $s$-wave pairing state may describe qualitatively
phonon-mediated superconductors. Our numerical findings reveal that the uniform $s$-wave pairing symmetry indeed produces an isotropic superconducting gap around the Fermi surface [Figs. ~\ref{fig.6} and ~\ref{fig.7}]. In addition to the uniform $s$-wave pairing, the NN-$s$-wave pairing also generates an isotropic pairing around the Fermi surface. 
However, for NN $s$-wave pairing, the energy gap around the Fermi surface is minimal, with $\Delta \approx 0.1\Delta_0$. Moreover, the superconducting critical temperature is generally determined by the gap magnitude $\Delta_0$. As such, for the NN $s$-wave pairing symmetry, we expect this ratio to be significantly smaller than the experimentally observed value. The other seven unconventional pairing states generate anisotropic energy gaps around the Fermi surface, as seen from Figs.~\ref{fig.6} and \ref{fig.7}, which deviate from the experimental results~\cite{Zhong2023}.

Recently, several different kagome-lattice-based superconductors have been reported, such as the ATi$_3$Bi$_5$ family ~\cite{yang2022titaniumbased,yang2022superconductivity,Li_2023}, the Ta$_2$V$_{3.1}$Si$_{0.9}$ material~\cite{liu2023vanadiumbased}, and the CsCr$_3$Sb$_5$ material~\cite{xu2023frustrated,liu2023superconductivity}. Some physical properties of these new kagome superconductors differ from those of AV$_3$Sb$_5$. The superconducting transition temperature of the ATi$_3$Bi$_5$ family is approximately 4.8 K, higher than the AV$_3$Sb$_5$ compounds (around 2.5 K). Additionally, this family of materials exhibits electronic nematicity and no charge-density-wave states~\cite{yang2022titaniumbased,yang2022superconductivity,Li_2023}. The Ta$_2$V$_{3.1}$Si$_{0.9}$ material has a much higher superconducting critical temperature, with T$_c$ at about 7.5 K~\cite{liu2023vanadiumbased}. Furthermore, it has been proposed that the CsCr$_3$Sb$_5$ material exhibits a strong correlation effect ~\cite{xu2023frustrated,liu2023superconductivity}. It is worth noting that no STM and ARPES experiments have been reported for these new kagome-based superconducting materials to date. Further studies may be required to explore the corresponding pairing symmetries and single-particle spectra.

\section{\label{sec4.sum}SUMMARY} In summary, we analyzed the single-particle spectra of kagome superconductors considering nine different pairing symmetries. The superconducting states with different pairing symmetries can be classified into irreducible representations of the $C_{6v}$ point group. Based on the numerical results of the energy bands and spectral function, three distinct pairing states were found, notably, the isotropic fully gapped states (uniform $s$-wave and NN $s$-wave), the anisotropic fully gapped states (NN $p+ip$ pairing, NN $d+id$ pairing, and NNN $s$-wave), and the nodal states (NN $f$-wave, NNN $p+ip$ pairing, NNN $d+id$, and NNN $f$-wave). The anisotropic fully gapped pairing states and the nodal pairing states can be further discerned by investigating the momentum-dependent superconducting gap and nodal points. The role of the impurity effect was also studied. For the $p+ip$ pairing and the $f$-wave pairing, sharp resonant peaks were revealed near an impurity. Whereas some in-gap features were observed for the $d+id$ symmetry, no in-gap features were observed for the three $s$-wave pairing states. These elements can be employed to probe the pairing symmetries of kagome superconductors.

To conclude, our work holds substantial significance in this field of study for three primary reasons. Firstly, there is no clear consensus presently among researchers regarding the pairing mechanism and symmetry of kagome superconductors; our work offers pivotal insights and knowledge that can be of value to experts working in this particular domain. Secondly, we have considered every relevant pairing symmetry, based on the spatial point group symmetry of the kagome lattice. Recently, new kagome superconductor compounds, including the ATi$_3$Bi$_5$ family~\cite{yang2022titaniumbased,yang2022superconductivity,Li_2023}, Ta$_2$V$_{3.1}$Si$_{0.9}$ material~\cite{liu2023vanadiumbased}, and CsCr$_3$Sb$_5$ material~\cite{xu2023frustrated,liu2023superconductivity}, have been reported.
 Our theoretical findings are not merely applicable to the early existing kagome superconductors but may extend to
these novel compounds, underlining the timeliness and relevancy of our research.
Lastly, for the first time, our study provides a systematic analysis of single-particle spectra, including the spectral function and the single impurity effect of kagome superconductors. Our findings highlight the potential use of single-particle spectra as an effective tool in determining the pairing symmetries of these type of superconductors, promising to facilitate a more comprehensive understanding of this rapidly evolving research area.
\begin{acknowledgments}
	This work was supported by the NSFC (Grant No.12074130) and the Natural Science Foundation of Guangdong Province (Grant No. 2021A1515012340).
\end{acknowledgments}


\begin{thebibliography}{52}%
\makeatletter
\providecommand \@ifxundefined [1]{%
 \@ifx{#1\undefined}
}%
\providecommand \@ifnum [1]{%
 \ifnum #1\expandafter \@firstoftwo
 \else \expandafter \@secondoftwo
 \fi
}%
\providecommand \@ifx [1]{%
 \ifx #1\expandafter \@firstoftwo
 \else \expandafter \@secondoftwo
 \fi
}%
\providecommand \natexlab [1]{#1}%
\providecommand \enquote  [1]{``#1''}%
\providecommand \bibnamefont  [1]{#1}%
\providecommand \bibfnamefont [1]{#1}%
\providecommand \citenamefont [1]{#1}%
\providecommand \href@noop [0]{\@secondoftwo}%
\providecommand \href [0]{\begingroup \@sanitize@url \@href}%
\providecommand \@href[1]{\@@startlink{#1}\@@href}%
\providecommand \@@href[1]{\endgroup#1\@@endlink}%
\providecommand \@sanitize@url [0]{\catcode `\\12\catcode `\$12\catcode
  `\&12\catcode `\#12\catcode `\^12\catcode `\_12\catcode `\%12\relax}%
\providecommand \@@startlink[1]{}%
\providecommand \@@endlink[0]{}%
\providecommand \url  [0]{\begingroup\@sanitize@url \@url }%
\providecommand \@url [1]{\endgroup\@href {#1}{\urlprefix }}%
\providecommand \urlprefix  [0]{URL }%
\providecommand \Eprint [0]{\href }%
\providecommand \doibase [0]{https://doi.org/}%
\providecommand \selectlanguage [0]{\@gobble}%
\providecommand \bibinfo  [0]{\@secondoftwo}%
\providecommand \bibfield  [0]{\@secondoftwo}%
\providecommand \translation [1]{[#1]}%
\providecommand \BibitemOpen [0]{}%
\providecommand \bibitemStop [0]{}%
\providecommand \bibitemNoStop [0]{.\EOS\space}%
\providecommand \EOS [0]{\spacefactor3000\relax}%
\providecommand \BibitemShut  [1]{\csname bibitem#1\endcsname}%
\let\auto@bib@innerbib\@empty
\bibitem [{\citenamefont {Kudo}\ \emph {et~al.}(2020)\citenamefont {Kudo},
  \citenamefont {Hiiragi}, \citenamefont {Honda}, \citenamefont {Fujimura},
  \citenamefont {Idei},\ and\ \citenamefont
  {Nohara}}]{doi:10.7566/JPSJ.89.013701}%
  \BibitemOpen
  \bibfield  {author} {\bibinfo {author} {\bibfnamefont {K.}~\bibnamefont
  {Kudo}}, \bibinfo {author} {\bibfnamefont {H.}~\bibnamefont {Hiiragi}},
  \bibinfo {author} {\bibfnamefont {T.}~\bibnamefont {Honda}}, \bibinfo
  {author} {\bibfnamefont {K.}~\bibnamefont {Fujimura}}, \bibinfo {author}
  {\bibfnamefont {H.}~\bibnamefont {Idei}},\ and\ \bibinfo {author}
  {\bibfnamefont {M.}~\bibnamefont {Nohara}},\ }\bibfield  {title} {\bibinfo
  {title} {Superconductivity in $\mathrm{Mg2Ir3Si}$: A fully ordered laves
  phase},\ }\href {https://doi.org/10.7566/JPSJ.89.013701} {\bibfield
  {journal} {\bibinfo  {journal} {J. Phys. Soc. Jpn.}\ }\textbf {\bibinfo
  {volume} {89}},\ \bibinfo {pages} {013701} (\bibinfo {year}
  {2020})}\BibitemShut {NoStop}%
\bibitem [{\citenamefont {Ku}\ \emph {et~al.}(1980)\citenamefont {Ku},
  \citenamefont {Meisner}, \citenamefont {Acker},\ and\ \citenamefont
  {Johnston}}]{KU198091}%
  \BibitemOpen
  \bibfield  {author} {\bibinfo {author} {\bibfnamefont {H.}~\bibnamefont
  {Ku}}, \bibinfo {author} {\bibfnamefont {G.}~\bibnamefont {Meisner}},
  \bibinfo {author} {\bibfnamefont {F.}~\bibnamefont {Acker}},\ and\ \bibinfo
  {author} {\bibfnamefont {D.}~\bibnamefont {Johnston}},\ }\bibfield  {title}
  {\bibinfo {title} {Superconducting and magnetic properties of new ternary
  borides with the $\mathrm{CeCo3B2}$-type structure},\ }\href
  {https://doi.org/https://doi.org/10.1016/0038-1098(80)90221-5} {\bibfield
  {journal} {\bibinfo  {journal} {Solid State Commun.}\ }\textbf {\bibinfo
  {volume} {35}},\ \bibinfo {pages} {91} (\bibinfo {year} {1980})}\BibitemShut
  {NoStop}%
\bibitem [{\citenamefont {Athreya}\ \emph {et~al.}(1985)\citenamefont
  {Athreya}, \citenamefont {Hausermann-Berg}, \citenamefont {Shelton},
  \citenamefont {Malik}, \citenamefont {Umarji},\ and\ \citenamefont
  {Shenoy}}]{ATHREYA1985330}%
  \BibitemOpen
  \bibfield  {author} {\bibinfo {author} {\bibfnamefont {K.}~\bibnamefont
  {Athreya}}, \bibinfo {author} {\bibfnamefont {L.}~\bibnamefont
  {Hausermann-Berg}}, \bibinfo {author} {\bibfnamefont {R.}~\bibnamefont
  {Shelton}}, \bibinfo {author} {\bibfnamefont {S.}~\bibnamefont {Malik}},
  \bibinfo {author} {\bibfnamefont {A.}~\bibnamefont {Umarji}},\ and\ \bibinfo
  {author} {\bibfnamefont {G.}~\bibnamefont {Shenoy}},\ }\bibfield  {title}
  {\bibinfo {title} {Superconductivity in the ternary borides
  $\mathrm{CeOs3B2}$ and $\mathrm{CeRu3B2}$: Magnetic susceptibility and
  specific heat measurements},\ }\href
  {https://doi.org/https://doi.org/10.1016/0375-9601(85)90177-X} {\bibfield
  {journal} {\bibinfo  {journal} {Phys. Lett. A}\ }\textbf {\bibinfo {volume}
  {113}},\ \bibinfo {pages} {330} (\bibinfo {year} {1985})}\BibitemShut
  {NoStop}%
\bibitem [{\citenamefont {Li}\ \emph {et~al.}(2011)\citenamefont {Li},
  \citenamefont {Zeng}, \citenamefont {Wan}, \citenamefont {Tao}, \citenamefont
  {Han}, \citenamefont {Yang}, \citenamefont {Wang},\ and\ \citenamefont
  {Wen}}]{PhysRevB.84.214527}%
  \BibitemOpen
  \bibfield  {author} {\bibinfo {author} {\bibfnamefont {S.}~\bibnamefont
  {Li}}, \bibinfo {author} {\bibfnamefont {B.}~\bibnamefont {Zeng}}, \bibinfo
  {author} {\bibfnamefont {X.}~\bibnamefont {Wan}}, \bibinfo {author}
  {\bibfnamefont {J.}~\bibnamefont {Tao}}, \bibinfo {author} {\bibfnamefont
  {F.}~\bibnamefont {Han}}, \bibinfo {author} {\bibfnamefont {H.}~\bibnamefont
  {Yang}}, \bibinfo {author} {\bibfnamefont {Z.}~\bibnamefont {Wang}},\ and\
  \bibinfo {author} {\bibfnamefont {H.-H.}\ \bibnamefont {Wen}},\ }\bibfield
  {title} {\bibinfo {title} {Anomalous properties in the normal and
  superconducting states of laru${}_{3}$si${}_{2}$},\ }\href
  {https://doi.org/10.1103/PhysRevB.84.214527} {\bibfield  {journal} {\bibinfo
  {journal} {Phys. Rev. B}\ }\textbf {\bibinfo {volume} {84}},\ \bibinfo
  {pages} {214527} (\bibinfo {year} {2011})}\BibitemShut {NoStop}%
\bibitem [{\citenamefont {Li}\ \emph {et~al.}(2015)\citenamefont {Li},
  \citenamefont {Xing}, \citenamefont {Tao}, \citenamefont {Yang},\ and\
  \citenamefont {Wen}}]{LI2015248}%
  \BibitemOpen
  \bibfield  {author} {\bibinfo {author} {\bibfnamefont {S.}~\bibnamefont
  {Li}}, \bibinfo {author} {\bibfnamefont {J.}~\bibnamefont {Xing}}, \bibinfo
  {author} {\bibfnamefont {J.}~\bibnamefont {Tao}}, \bibinfo {author}
  {\bibfnamefont {H.}~\bibnamefont {Yang}},\ and\ \bibinfo {author}
  {\bibfnamefont {H.-H.}\ \bibnamefont {Wen}},\ }\bibfield  {title} {\bibinfo
  {title} {Superconductivity in $\mathrm{Ba2/3Pt3B2}$ with the kagome
  lattice},\ }\href {https://doi.org/https://doi.org/10.1016/j.aop.2015.04.008}
  {\bibfield  {journal} {\bibinfo  {journal} {Ann. Phys.}\ }\textbf {\bibinfo
  {volume} {358}},\ \bibinfo {pages} {248} (\bibinfo {year}
  {2015})}\BibitemShut {NoStop}%
\bibitem [{\citenamefont {Chaudhary}\ \emph {et~al.}(2023)\citenamefont
  {Chaudhary}, \citenamefont {Shama}, \citenamefont {Singh}, \citenamefont
  {Consiglio}, \citenamefont {Di~Sante}, \citenamefont {Thomale},\ and\
  \citenamefont {Singh}}]{PhysRevB.107.085103}%
  \BibitemOpen
  \bibfield  {author} {\bibinfo {author} {\bibfnamefont {S.}~\bibnamefont
  {Chaudhary}}, \bibinfo {author} {\bibnamefont {Shama}}, \bibinfo {author}
  {\bibfnamefont {J.}~\bibnamefont {Singh}}, \bibinfo {author} {\bibfnamefont
  {A.}~\bibnamefont {Consiglio}}, \bibinfo {author} {\bibfnamefont
  {D.}~\bibnamefont {Di~Sante}}, \bibinfo {author} {\bibfnamefont
  {R.}~\bibnamefont {Thomale}},\ and\ \bibinfo {author} {\bibfnamefont
  {Y.}~\bibnamefont {Singh}},\ }\bibfield  {title} {\bibinfo {title} {Role of
  electronic correlations in the kagome-lattice superconductor
  $\mathrm{LaRh}_{3}\mathrm{B}_{2}$},\ }\href
  {https://doi.org/10.1103/PhysRevB.107.085103} {\bibfield  {journal} {\bibinfo
   {journal} {Phys. Rev. B}\ }\textbf {\bibinfo {volume} {107}},\ \bibinfo
  {pages} {085103} (\bibinfo {year} {2023})}\BibitemShut {NoStop}%
\bibitem [{\citenamefont {Gui}\ and\ \citenamefont {Cava}(2022)}]{Gui2022}%
  \BibitemOpen
  \bibfield  {author} {\bibinfo {author} {\bibfnamefont {X.}~\bibnamefont
  {Gui}}\ and\ \bibinfo {author} {\bibfnamefont {R.~J.}\ \bibnamefont {Cava}},\
  }\bibfield  {title} {\bibinfo {title} {$\mathrm{LaIr3Ga2}$: A superconductor
  based on a kagome lattice of $\mathrm{Ir}$},\ }\href
  {https://doi.org/10.1021/acs.chemmater.2c00280} {\bibfield  {journal}
  {\bibinfo  {journal} {Chemistry of Materials}\ }\textbf {\bibinfo {volume}
  {34}},\ \bibinfo {pages} {2824} (\bibinfo {year} {2022})}\BibitemShut
  {NoStop}%
\bibitem [{\citenamefont {Gong}\ \emph {et~al.}(2022)\citenamefont {Gong},
  \citenamefont {Tian}, \citenamefont {Tu}, \citenamefont {Yin}, \citenamefont
  {Fu}, \citenamefont {Luo},\ and\ \citenamefont {Lei}}]{Gong_2022}%
  \BibitemOpen
  \bibfield  {author} {\bibinfo {author} {\bibfnamefont {C.}~\bibnamefont
  {Gong}}, \bibinfo {author} {\bibfnamefont {S.}~\bibnamefont {Tian}}, \bibinfo
  {author} {\bibfnamefont {Z.}~\bibnamefont {Tu}}, \bibinfo {author}
  {\bibfnamefont {Q.}~\bibnamefont {Yin}}, \bibinfo {author} {\bibfnamefont
  {Y.}~\bibnamefont {Fu}}, \bibinfo {author} {\bibfnamefont {R.}~\bibnamefont
  {Luo}},\ and\ \bibinfo {author} {\bibfnamefont {H.}~\bibnamefont {Lei}},\
  }\bibfield  {title} {\bibinfo {title} {Superconductivity in kagome metal
  $\mathrm{YRu3Si2}$ with strong electron correlations},\ }\href
  {https://doi.org/10.1088/0256-307X/39/8/087401} {\bibfield  {journal}
  {\bibinfo  {journal} {Chinese Physics Letters}\ }\textbf {\bibinfo {volume}
  {39}},\ \bibinfo {pages} {087401} (\bibinfo {year} {2022})}\BibitemShut
  {NoStop}%
\bibitem [{\citenamefont {Ortiz}\ \emph {et~al.}(2020)\citenamefont {Ortiz},
  \citenamefont {Teicher}, \citenamefont {Hu}, \citenamefont {Zuo},
  \citenamefont {Sarte}, \citenamefont {Schueller}, \citenamefont {Abeykoon},
  \citenamefont {Krogstad}, \citenamefont {Rosenkranz}, \citenamefont {Osborn},
  \citenamefont {Seshadri}, \citenamefont {Balents}, \citenamefont {He},\ and\
  \citenamefont {Wilson}}]{r11}%
  \BibitemOpen
  \bibfield  {author} {\bibinfo {author} {\bibfnamefont {B.~R.}\ \bibnamefont
  {Ortiz}}, \bibinfo {author} {\bibfnamefont {S.~M.~L.}\ \bibnamefont
  {Teicher}}, \bibinfo {author} {\bibfnamefont {Y.}~\bibnamefont {Hu}},
  \bibinfo {author} {\bibfnamefont {J.~L.}\ \bibnamefont {Zuo}}, \bibinfo
  {author} {\bibfnamefont {P.~M.}\ \bibnamefont {Sarte}}, \bibinfo {author}
  {\bibfnamefont {E.~C.}\ \bibnamefont {Schueller}}, \bibinfo {author}
  {\bibfnamefont {A.~M.~M.}\ \bibnamefont {Abeykoon}}, \bibinfo {author}
  {\bibfnamefont {M.~J.}\ \bibnamefont {Krogstad}}, \bibinfo {author}
  {\bibfnamefont {S.}~\bibnamefont {Rosenkranz}}, \bibinfo {author}
  {\bibfnamefont {R.}~\bibnamefont {Osborn}}, \bibinfo {author} {\bibfnamefont
  {R.}~\bibnamefont {Seshadri}}, \bibinfo {author} {\bibfnamefont
  {L.}~\bibnamefont {Balents}}, \bibinfo {author} {\bibfnamefont
  {J.}~\bibnamefont {He}},\ and\ \bibinfo {author} {\bibfnamefont {S.~D.}\
  \bibnamefont {Wilson}},\ }\bibfield  {title} {\bibinfo {title}
  {{$\mathrm{Cs}{\mathrm{V}}_{3}{\mathrm{Sb}}_{5}$}: A {${\mathbb{Z}}_{2}$}
  topological kagome metal with a superconducting ground state},\ }\href
  {https://doi.org/10.1103/PhysRevLett.125.247002} {\bibfield  {journal}
  {\bibinfo  {journal} {Phys. Rev. Lett.}\ }\textbf {\bibinfo {volume} {125}},\
  \bibinfo {pages} {247002} (\bibinfo {year} {2020})}\BibitemShut {NoStop}%
\bibitem [{\citenamefont {Yin}\ \emph {et~al.}(2021)\citenamefont {Yin},
  \citenamefont {Tu}, \citenamefont {Gong},\ and\ \citenamefont {Lei}}]{r12}%
  \BibitemOpen
  \bibfield  {author} {\bibinfo {author} {\bibfnamefont {Q.}~\bibnamefont
  {Yin}}, \bibinfo {author} {\bibfnamefont {Z.}~\bibnamefont {Tu}}, \bibinfo
  {author} {\bibfnamefont {C.}~\bibnamefont {Gong}},\ and\ \bibinfo {author}
  {\bibfnamefont {Y.~F. S. Y.~H.}\ \bibnamefont {Lei}},\ }\bibfield  {title}
  {\bibinfo {title} {Superconductivity and normal-state properties of kagome
  metal {${\mathrm{RbV}}_{3}{\mathrm{Sb}}_{5}$} single crystals},\ }\href
  {https://doi.org/10.1088/0256-307X/38/3/037403} {\bibfield  {journal}
  {\bibinfo  {journal} {Chin. Phys. Lett.}\ }\textbf {\bibinfo {volume} {38}},\
  \bibinfo {eid} {037403} (\bibinfo {year} {2021})}\BibitemShut {NoStop}%
\bibitem [{\citenamefont {Ortiz}\ \emph {et~al.}(2021)\citenamefont {Ortiz},
  \citenamefont {Sarte}, \citenamefont {Kenney}, \citenamefont {Graf},
  \citenamefont {Teicher}, \citenamefont {Seshadri},\ and\ \citenamefont
  {Wilson}}]{r13}%
  \BibitemOpen
  \bibfield  {author} {\bibinfo {author} {\bibfnamefont {B.~R.}\ \bibnamefont
  {Ortiz}}, \bibinfo {author} {\bibfnamefont {P.~M.}\ \bibnamefont {Sarte}},
  \bibinfo {author} {\bibfnamefont {E.~M.}\ \bibnamefont {Kenney}}, \bibinfo
  {author} {\bibfnamefont {M.~J.}\ \bibnamefont {Graf}}, \bibinfo {author}
  {\bibfnamefont {S.~M.~L.}\ \bibnamefont {Teicher}}, \bibinfo {author}
  {\bibfnamefont {R.}~\bibnamefont {Seshadri}},\ and\ \bibinfo {author}
  {\bibfnamefont {S.~D.}\ \bibnamefont {Wilson}},\ }\bibfield  {title}
  {\bibinfo {title} {Superconductivity in the {${\mathbb{Z}}_{2}$} kagome metal
  {${\mathrm{KV}}_{3}{\mathrm{Sb}}_{5}$}},\ }\href
  {https://doi.org/10.1103/PhysRevMaterials.5.034801} {\bibfield  {journal}
  {\bibinfo  {journal} {Phys. Rev. Mater.}\ }\textbf {\bibinfo {volume} {5}},\
  \bibinfo {pages} {034801} (\bibinfo {year} {2021})}\BibitemShut {NoStop}%
\bibitem [{\citenamefont {Yang}\ \emph
  {et~al.}(2022{\natexlab{a}})\citenamefont {Yang}, \citenamefont {Zhao},
  \citenamefont {Yi}, \citenamefont {Liu}, \citenamefont {You}, \citenamefont
  {Zhang}, \citenamefont {Guo}, \citenamefont {Lin}, \citenamefont {Shen},
  \citenamefont {Chen}, \citenamefont {Dong}, \citenamefont {Su},\ and\
  \citenamefont {Gao}}]{yang2022titaniumbased}%
  \BibitemOpen
  \bibfield  {author} {\bibinfo {author} {\bibfnamefont {H.}~\bibnamefont
  {Yang}}, \bibinfo {author} {\bibfnamefont {Z.}~\bibnamefont {Zhao}}, \bibinfo
  {author} {\bibfnamefont {X.-W.}\ \bibnamefont {Yi}}, \bibinfo {author}
  {\bibfnamefont {J.}~\bibnamefont {Liu}}, \bibinfo {author} {\bibfnamefont
  {J.-Y.}\ \bibnamefont {You}}, \bibinfo {author} {\bibfnamefont
  {Y.}~\bibnamefont {Zhang}}, \bibinfo {author} {\bibfnamefont
  {H.}~\bibnamefont {Guo}}, \bibinfo {author} {\bibfnamefont {X.}~\bibnamefont
  {Lin}}, \bibinfo {author} {\bibfnamefont {C.}~\bibnamefont {Shen}}, \bibinfo
  {author} {\bibfnamefont {H.}~\bibnamefont {Chen}}, \bibinfo {author}
  {\bibfnamefont {X.}~\bibnamefont {Dong}}, \bibinfo {author} {\bibfnamefont
  {G.}~\bibnamefont {Su}},\ and\ \bibinfo {author} {\bibfnamefont {H.-J.}\
  \bibnamefont {Gao}},\ }\href@noop {} {\bibinfo {title} {Titanium-based kagome
  superconductor csti$_3$bi$_5$ and topological states}} (\bibinfo {year}
  {2022}{\natexlab{a}}),\ \Eprint {https://arxiv.org/abs/2209.03840}
  {arXiv:2209.03840 [cond-mat.supr-con]} \BibitemShut {NoStop}%
\bibitem [{\citenamefont {Yang}\ \emph
  {et~al.}(2022{\natexlab{b}})\citenamefont {Yang}, \citenamefont {Ye},
  \citenamefont {Zhao}, \citenamefont {Liu}, \citenamefont {Yi}, \citenamefont
  {Zhang}, \citenamefont {Shi}, \citenamefont {You}, \citenamefont {Huang},
  \citenamefont {Wang}, \citenamefont {Wang}, \citenamefont {Guo},
  \citenamefont {Lin}, \citenamefont {Shen}, \citenamefont {Zhou},
  \citenamefont {Chen}, \citenamefont {Dong}, \citenamefont {Su}, \citenamefont
  {Wang},\ and\ \citenamefont {Gao}}]{yang2022superconductivity}%
  \BibitemOpen
  \bibfield  {author} {\bibinfo {author} {\bibfnamefont {H.}~\bibnamefont
  {Yang}}, \bibinfo {author} {\bibfnamefont {Y.}~\bibnamefont {Ye}}, \bibinfo
  {author} {\bibfnamefont {Z.}~\bibnamefont {Zhao}}, \bibinfo {author}
  {\bibfnamefont {J.}~\bibnamefont {Liu}}, \bibinfo {author} {\bibfnamefont
  {X.-W.}\ \bibnamefont {Yi}}, \bibinfo {author} {\bibfnamefont
  {Y.}~\bibnamefont {Zhang}}, \bibinfo {author} {\bibfnamefont
  {J.}~\bibnamefont {Shi}}, \bibinfo {author} {\bibfnamefont {J.-Y.}\
  \bibnamefont {You}}, \bibinfo {author} {\bibfnamefont {Z.}~\bibnamefont
  {Huang}}, \bibinfo {author} {\bibfnamefont {B.}~\bibnamefont {Wang}},
  \bibinfo {author} {\bibfnamefont {J.}~\bibnamefont {Wang}}, \bibinfo {author}
  {\bibfnamefont {H.}~\bibnamefont {Guo}}, \bibinfo {author} {\bibfnamefont
  {X.}~\bibnamefont {Lin}}, \bibinfo {author} {\bibfnamefont {C.}~\bibnamefont
  {Shen}}, \bibinfo {author} {\bibfnamefont {W.}~\bibnamefont {Zhou}}, \bibinfo
  {author} {\bibfnamefont {H.}~\bibnamefont {Chen}}, \bibinfo {author}
  {\bibfnamefont {X.}~\bibnamefont {Dong}}, \bibinfo {author} {\bibfnamefont
  {G.}~\bibnamefont {Su}}, \bibinfo {author} {\bibfnamefont {Z.}~\bibnamefont
  {Wang}},\ and\ \bibinfo {author} {\bibfnamefont {H.-J.}\ \bibnamefont
  {Gao}},\ }\href@noop {} {\bibinfo {title} {Superconductivity and
  orbital-selective nematic order in a new titanium-based kagome metal
  csti3bi5}} (\bibinfo {year} {2022}{\natexlab{b}}),\ \Eprint
  {https://arxiv.org/abs/2211.12264} {arXiv:2211.12264 [cond-mat.supr-con]}
  \BibitemShut {NoStop}%
\bibitem [{\citenamefont {Li}\ \emph {et~al.}(2023)\citenamefont {Li},
  \citenamefont {Cheng}, \citenamefont {Ortiz}, \citenamefont {Tan},
  \citenamefont {Werhahn}, \citenamefont {Zeng}, \citenamefont {Johrendt},
  \citenamefont {Yan}, \citenamefont {Wang}, \citenamefont {Wilson},\ and\
  \citenamefont {Zeljkovic}}]{Li_2023}%
  \BibitemOpen
  \bibfield  {author} {\bibinfo {author} {\bibfnamefont {H.}~\bibnamefont
  {Li}}, \bibinfo {author} {\bibfnamefont {S.}~\bibnamefont {Cheng}}, \bibinfo
  {author} {\bibfnamefont {B.~R.}\ \bibnamefont {Ortiz}}, \bibinfo {author}
  {\bibfnamefont {H.}~\bibnamefont {Tan}}, \bibinfo {author} {\bibfnamefont
  {D.}~\bibnamefont {Werhahn}}, \bibinfo {author} {\bibfnamefont
  {K.}~\bibnamefont {Zeng}}, \bibinfo {author} {\bibfnamefont {D.}~\bibnamefont
  {Johrendt}}, \bibinfo {author} {\bibfnamefont {B.}~\bibnamefont {Yan}},
  \bibinfo {author} {\bibfnamefont {Z.}~\bibnamefont {Wang}}, \bibinfo {author}
  {\bibfnamefont {S.~D.}\ \bibnamefont {Wilson}},\ and\ \bibinfo {author}
  {\bibfnamefont {I.}~\bibnamefont {Zeljkovic}},\ }\bibfield  {title} {\bibinfo
  {title} {Electronic nematicity without charge density waves in titanium-based
  kagome metal},\ }\bibfield  {journal} {\bibinfo  {journal} {Nat. Phys.}\
  }\href {https://doi.org/10.1038/s41567-023-02176-3}
  {10.1038/s41567-023-02176-3} (\bibinfo {year} {2023})\BibitemShut {NoStop}%
\bibitem [{\citenamefont {Liu}\ \emph {et~al.}(2023{\natexlab{a}})\citenamefont
  {Liu}, \citenamefont {Yao}, \citenamefont {Shi}, \citenamefont {Yang},
  \citenamefont {Yan}, \citenamefont {Li}, \citenamefont {Chen}, \citenamefont
  {Feng}, \citenamefont {Li}, \citenamefont {Wang},\ and\ \citenamefont
  {Shi}}]{liu2023vanadiumbased}%
  \BibitemOpen
  \bibfield  {author} {\bibinfo {author} {\bibfnamefont {H.}~\bibnamefont
  {Liu}}, \bibinfo {author} {\bibfnamefont {J.}~\bibnamefont {Yao}}, \bibinfo
  {author} {\bibfnamefont {J.}~\bibnamefont {Shi}}, \bibinfo {author}
  {\bibfnamefont {Z.}~\bibnamefont {Yang}}, \bibinfo {author} {\bibfnamefont
  {D.}~\bibnamefont {Yan}}, \bibinfo {author} {\bibfnamefont {Y.}~\bibnamefont
  {Li}}, \bibinfo {author} {\bibfnamefont {D.}~\bibnamefont {Chen}}, \bibinfo
  {author} {\bibfnamefont {H.~L.}\ \bibnamefont {Feng}}, \bibinfo {author}
  {\bibfnamefont {S.}~\bibnamefont {Li}}, \bibinfo {author} {\bibfnamefont
  {Z.}~\bibnamefont {Wang}},\ and\ \bibinfo {author} {\bibfnamefont
  {Y.}~\bibnamefont {Shi}},\ }\href@noop {} {\bibinfo {title} {Vanadium-based
  superconductivity in a breathing kagome compound $\mathrm{Ta2V3.1Si0.9}$}}
  (\bibinfo {year} {2023}{\natexlab{a}}),\ \Eprint
  {https://arxiv.org/abs/2306.03370} {arXiv:2306.03370 [cond-mat.supr-con]}
  \BibitemShut {NoStop}%
\bibitem [{\citenamefont {Xu}\ \emph {et~al.}(2023)\citenamefont {Xu},
  \citenamefont {Wu}, \citenamefont {Zhi}, \citenamefont {Cao}, \citenamefont
  {Dai}, \citenamefont {Cao}, \citenamefont {Wang},\ and\ \citenamefont
  {Lin}}]{xu2023frustrated}%
  \BibitemOpen
  \bibfield  {author} {\bibinfo {author} {\bibfnamefont {C.}~\bibnamefont
  {Xu}}, \bibinfo {author} {\bibfnamefont {S.}~\bibnamefont {Wu}}, \bibinfo
  {author} {\bibfnamefont {G.-X.}\ \bibnamefont {Zhi}}, \bibinfo {author}
  {\bibfnamefont {G.}~\bibnamefont {Cao}}, \bibinfo {author} {\bibfnamefont
  {J.}~\bibnamefont {Dai}}, \bibinfo {author} {\bibfnamefont {C.}~\bibnamefont
  {Cao}}, \bibinfo {author} {\bibfnamefont {X.}~\bibnamefont {Wang}},\ and\
  \bibinfo {author} {\bibfnamefont {H.-Q.}\ \bibnamefont {Lin}},\ }\href@noop
  {} {\bibinfo {title} {Frustrated altermagnetism and charge density wave in
  kagome superconductor cscr3sb5}} (\bibinfo {year} {2023}),\ \Eprint
  {https://arxiv.org/abs/2309.14812} {arXiv:2309.14812 [cond-mat.supr-con]}
  \BibitemShut {NoStop}%
\bibitem [{\citenamefont {Liu}\ \emph {et~al.}(2023{\natexlab{b}})\citenamefont
  {Liu}, \citenamefont {Liu}, \citenamefont {Bao}, \citenamefont {Yang},
  \citenamefont {Ji}, \citenamefont {Liu}, \citenamefont {Xu}, \citenamefont
  {Yang}, \citenamefont {Chai}, \citenamefont {Lu}, \citenamefont {Liu},
  \citenamefont {Wang}, \citenamefont {Jiang}, \citenamefont {Tao},
  \citenamefont {Ren}, \citenamefont {Xu}, \citenamefont {Cao}, \citenamefont
  {Xu}, \citenamefont {Cheng},\ and\ \citenamefont
  {Cao}}]{liu2023superconductivity}%
  \BibitemOpen
  \bibfield  {author} {\bibinfo {author} {\bibfnamefont {Y.}~\bibnamefont
  {Liu}}, \bibinfo {author} {\bibfnamefont {Z.-Y.}\ \bibnamefont {Liu}},
  \bibinfo {author} {\bibfnamefont {J.-K.}\ \bibnamefont {Bao}}, \bibinfo
  {author} {\bibfnamefont {P.-T.}\ \bibnamefont {Yang}}, \bibinfo {author}
  {\bibfnamefont {L.-W.}\ \bibnamefont {Ji}}, \bibinfo {author} {\bibfnamefont
  {J.-Y.}\ \bibnamefont {Liu}}, \bibinfo {author} {\bibfnamefont {C.-C.}\
  \bibnamefont {Xu}}, \bibinfo {author} {\bibfnamefont {W.-Z.}\ \bibnamefont
  {Yang}}, \bibinfo {author} {\bibfnamefont {W.-L.}\ \bibnamefont {Chai}},
  \bibinfo {author} {\bibfnamefont {J.-Y.}\ \bibnamefont {Lu}}, \bibinfo
  {author} {\bibfnamefont {C.-C.}\ \bibnamefont {Liu}}, \bibinfo {author}
  {\bibfnamefont {B.-S.}\ \bibnamefont {Wang}}, \bibinfo {author}
  {\bibfnamefont {H.}~\bibnamefont {Jiang}}, \bibinfo {author} {\bibfnamefont
  {Q.}~\bibnamefont {Tao}}, \bibinfo {author} {\bibfnamefont {Z.}~\bibnamefont
  {Ren}}, \bibinfo {author} {\bibfnamefont {X.-F.}\ \bibnamefont {Xu}},
  \bibinfo {author} {\bibfnamefont {C.}~\bibnamefont {Cao}}, \bibinfo {author}
  {\bibfnamefont {Z.-A.}\ \bibnamefont {Xu}}, \bibinfo {author} {\bibfnamefont
  {J.-G.}\ \bibnamefont {Cheng}},\ and\ \bibinfo {author} {\bibfnamefont
  {G.-H.}\ \bibnamefont {Cao}},\ }\href@noop {} {\bibinfo {title}
  {Superconductivity emerged from density-wave order in a kagome bad metal}}
  (\bibinfo {year} {2023}{\natexlab{b}}),\ \Eprint
  {https://arxiv.org/abs/2309.13514} {arXiv:2309.13514 [cond-mat.supr-con]}
  \BibitemShut {NoStop}%
\bibitem [{\citenamefont {Zhu}\ \emph {et~al.}(2022)\citenamefont {Zhu},
  \citenamefont {Yang}, \citenamefont {Xia}, \citenamefont {Yin}, \citenamefont
  {Wang}, \citenamefont {Zhao}, \citenamefont {Dai}, \citenamefont {Tu},
  \citenamefont {Song}, \citenamefont {Tao}, \citenamefont {Tu}, \citenamefont
  {Gong}, \citenamefont {Lei}, \citenamefont {Guo},\ and\ \citenamefont
  {Li}}]{r14}%
  \BibitemOpen
  \bibfield  {author} {\bibinfo {author} {\bibfnamefont {C.~C.}\ \bibnamefont
  {Zhu}}, \bibinfo {author} {\bibfnamefont {X.~F.}\ \bibnamefont {Yang}},
  \bibinfo {author} {\bibfnamefont {W.}~\bibnamefont {Xia}}, \bibinfo {author}
  {\bibfnamefont {Q.~W.}\ \bibnamefont {Yin}}, \bibinfo {author} {\bibfnamefont
  {L.~S.}\ \bibnamefont {Wang}}, \bibinfo {author} {\bibfnamefont {C.~C.}\
  \bibnamefont {Zhao}}, \bibinfo {author} {\bibfnamefont {D.~Z.}\ \bibnamefont
  {Dai}}, \bibinfo {author} {\bibfnamefont {C.~P.}\ \bibnamefont {Tu}},
  \bibinfo {author} {\bibfnamefont {B.~Q.}\ \bibnamefont {Song}}, \bibinfo
  {author} {\bibfnamefont {Z.~C.}\ \bibnamefont {Tao}}, \bibinfo {author}
  {\bibfnamefont {Z.~J.}\ \bibnamefont {Tu}}, \bibinfo {author} {\bibfnamefont
  {C.~S.}\ \bibnamefont {Gong}}, \bibinfo {author} {\bibfnamefont {H.~C.}\
  \bibnamefont {Lei}}, \bibinfo {author} {\bibfnamefont {Y.~F.}\ \bibnamefont
  {Guo}},\ and\ \bibinfo {author} {\bibfnamefont {S.~Y.}\ \bibnamefont {Li}},\
  }\bibfield  {title} {\bibinfo {title} {Double-dome superconductivity under
  pressure in the {V-based} kagome metals
  {$A{\mathrm{V}}_{3}{\mathrm{Sb}}_{5}$} ({$A=\mathrm{Rb}$ and K})},\ }\href
  {https://doi.org/10.1103/PhysRevB.105.094507} {\bibfield  {journal} {\bibinfo
   {journal} {Phys. Rev. B}\ }\textbf {\bibinfo {volume} {105}},\ \bibinfo
  {pages} {094507} (\bibinfo {year} {2022})}\BibitemShut {NoStop}%
\bibitem [{\citenamefont {Zhao}\ \emph {et~al.}(2021)\citenamefont {Zhao},
  \citenamefont {Wang}, \citenamefont {Xia}, \citenamefont {Yin}, \citenamefont
  {Ni}, \citenamefont {Huang}, \citenamefont {Tu}, \citenamefont {Tao},
  \citenamefont {Tu}, \citenamefont {Gong}, \citenamefont {Lei}, \citenamefont
  {Guo}, \citenamefont {Yang},\ and\ \citenamefont {Li}}]{r15}%
  \BibitemOpen
  \bibfield  {author} {\bibinfo {author} {\bibfnamefont {C.~C.}\ \bibnamefont
  {Zhao}}, \bibinfo {author} {\bibfnamefont {L.~S.}\ \bibnamefont {Wang}},
  \bibinfo {author} {\bibfnamefont {W.}~\bibnamefont {Xia}}, \bibinfo {author}
  {\bibfnamefont {Q.~W.}\ \bibnamefont {Yin}}, \bibinfo {author} {\bibfnamefont
  {J.~M.}\ \bibnamefont {Ni}}, \bibinfo {author} {\bibfnamefont {Y.~Y.}\
  \bibnamefont {Huang}}, \bibinfo {author} {\bibfnamefont {C.~P.}\ \bibnamefont
  {Tu}}, \bibinfo {author} {\bibfnamefont {Z.~C.}\ \bibnamefont {Tao}},
  \bibinfo {author} {\bibfnamefont {Z.~J.}\ \bibnamefont {Tu}}, \bibinfo
  {author} {\bibfnamefont {C.~S.}\ \bibnamefont {Gong}}, \bibinfo {author}
  {\bibfnamefont {H.~C.}\ \bibnamefont {Lei}}, \bibinfo {author} {\bibfnamefont
  {Y.~F.}\ \bibnamefont {Guo}}, \bibinfo {author} {\bibfnamefont {X.~F.}\
  \bibnamefont {Yang}},\ and\ \bibinfo {author} {\bibfnamefont {S.~Y.}\
  \bibnamefont {Li}},\ }\href@noop {} {\bibinfo {title} {Nodal
  superconductivity and superconducting domes in the topological kagome metal
  {${\mathrm{CsV}}_{3}{\mathrm{Sb}}_{5}$}}} (\bibinfo {year} {2021}),\ \Eprint
  {https://arxiv.org/abs/2102.08356} {arXiv:2102.08356} \BibitemShut {NoStop}%
\bibitem [{\citenamefont {Zhang}\ \emph {et~al.}(2021)\citenamefont {Zhang},
  \citenamefont {Chen}, \citenamefont {Zhou}, \citenamefont {Yuan},
  \citenamefont {Wang}, \citenamefont {Wang}, \citenamefont {Yang},
  \citenamefont {An}, \citenamefont {Zhang}, \citenamefont {Zhu}, \citenamefont
  {Zhou}, \citenamefont {Chen}, \citenamefont {Zhou},\ and\ \citenamefont
  {Yang}}]{r16}%
  \BibitemOpen
  \bibfield  {author} {\bibinfo {author} {\bibfnamefont {Z.}~\bibnamefont
  {Zhang}}, \bibinfo {author} {\bibfnamefont {Z.}~\bibnamefont {Chen}},
  \bibinfo {author} {\bibfnamefont {Y.}~\bibnamefont {Zhou}}, \bibinfo {author}
  {\bibfnamefont {Y.}~\bibnamefont {Yuan}}, \bibinfo {author} {\bibfnamefont
  {S.}~\bibnamefont {Wang}}, \bibinfo {author} {\bibfnamefont {J.}~\bibnamefont
  {Wang}}, \bibinfo {author} {\bibfnamefont {H.}~\bibnamefont {Yang}}, \bibinfo
  {author} {\bibfnamefont {C.}~\bibnamefont {An}}, \bibinfo {author}
  {\bibfnamefont {L.}~\bibnamefont {Zhang}}, \bibinfo {author} {\bibfnamefont
  {X.}~\bibnamefont {Zhu}}, \bibinfo {author} {\bibfnamefont {Y.}~\bibnamefont
  {Zhou}}, \bibinfo {author} {\bibfnamefont {X.}~\bibnamefont {Chen}}, \bibinfo
  {author} {\bibfnamefont {J.}~\bibnamefont {Zhou}},\ and\ \bibinfo {author}
  {\bibfnamefont {Z.}~\bibnamefont {Yang}},\ }\bibfield  {title} {\bibinfo
  {title} {Pressure-induced reemergence of superconductivity in the topological
  kagome metal {$\mathrm{Cs}{\mathrm{V}}_{3}{\mathrm{Sb}}_{5}$}},\ }\href
  {https://doi.org/10.1103/PhysRevB.103.224513} {\bibfield  {journal} {\bibinfo
   {journal} {Phys. Rev. B}\ }\textbf {\bibinfo {volume} {103}},\ \bibinfo
  {pages} {224513} (\bibinfo {year} {2021})}\BibitemShut {NoStop}%
\bibitem [{\citenamefont {Chen}\ \emph {et~al.}(2021)\citenamefont {Chen},
  \citenamefont {Zhan}, \citenamefont {Wang}, \citenamefont {Deng},
  \citenamefont {Liu}, \citenamefont {Chen}, \citenamefont {Guo},\ and\
  \citenamefont {Chen}}]{r17}%
  \BibitemOpen
  \bibfield  {author} {\bibinfo {author} {\bibfnamefont {X.}~\bibnamefont
  {Chen}}, \bibinfo {author} {\bibfnamefont {X.}~\bibnamefont {Zhan}}, \bibinfo
  {author} {\bibfnamefont {X.}~\bibnamefont {Wang}}, \bibinfo {author}
  {\bibfnamefont {J.}~\bibnamefont {Deng}}, \bibinfo {author} {\bibfnamefont
  {X.-B.}\ \bibnamefont {Liu}}, \bibinfo {author} {\bibfnamefont
  {X.}~\bibnamefont {Chen}}, \bibinfo {author} {\bibfnamefont {J.-G.}\
  \bibnamefont {Guo}},\ and\ \bibinfo {author} {\bibfnamefont {X.}~\bibnamefont
  {Chen}},\ }\bibfield  {title} {\bibinfo {title} {Highly robust reentrant
  superconductivity in {${\mathrm{CsV}}_{3}{\mathrm{Sb}}_{5}$} under
  pressure},\ }\href {https://doi.org/10.1088/0256-307x/38/5/057402} {\bibfield
   {journal} {\bibinfo  {journal} {Chin. Phys. Lett.}\ }\textbf {\bibinfo
  {volume} {38}},\ \bibinfo {pages} {057402} (\bibinfo {year}
  {2021})}\BibitemShut {NoStop}%
\bibitem [{\citenamefont {Yu}\ \emph {et~al.}(2022)\citenamefont {Yu},
  \citenamefont {Zhu}, \citenamefont {Wen}, \citenamefont {Gui}, \citenamefont
  {Li}, \citenamefont {Han}, \citenamefont {Wu}, \citenamefont {Wang},
  \citenamefont {Xiang}, \citenamefont {Qiao}, \citenamefont {Ying},\ and\
  \citenamefont {Chen}}]{r18}%
  \BibitemOpen
  \bibfield  {author} {\bibinfo {author} {\bibfnamefont {F.}~\bibnamefont
  {Yu}}, \bibinfo {author} {\bibfnamefont {X.}~\bibnamefont {Zhu}}, \bibinfo
  {author} {\bibfnamefont {X.}~\bibnamefont {Wen}}, \bibinfo {author}
  {\bibfnamefont {Z.}~\bibnamefont {Gui}}, \bibinfo {author} {\bibfnamefont
  {Z.}~\bibnamefont {Li}}, \bibinfo {author} {\bibfnamefont {Y.}~\bibnamefont
  {Han}}, \bibinfo {author} {\bibfnamefont {T.}~\bibnamefont {Wu}}, \bibinfo
  {author} {\bibfnamefont {Z.}~\bibnamefont {Wang}}, \bibinfo {author}
  {\bibfnamefont {Z.}~\bibnamefont {Xiang}}, \bibinfo {author} {\bibfnamefont
  {Z.}~\bibnamefont {Qiao}}, \bibinfo {author} {\bibfnamefont {J.}~\bibnamefont
  {Ying}},\ and\ \bibinfo {author} {\bibfnamefont {X.}~\bibnamefont {Chen}},\
  }\bibfield  {title} {\bibinfo {title} {Pressure-induced dimensional crossover
  in a kagome superconductor},\ }\href
  {https://doi.org/10.1103/PhysRevLett.128.077001} {\bibfield  {journal}
  {\bibinfo  {journal} {Phys. Rev. Lett.}\ }\textbf {\bibinfo {volume} {128}},\
  \bibinfo {pages} {077001} (\bibinfo {year} {2022})}\BibitemShut {NoStop}%
\bibitem [{\citenamefont {Yang}\ \emph {et~al.}(2020)\citenamefont {Yang},
  \citenamefont {Wang}, \citenamefont {Ortiz}, \citenamefont {Liu},
  \citenamefont {Gayles}, \citenamefont {Derunova}, \citenamefont
  {Gonzalez-Hernandez}, \citenamefont {艩mejkal}, \citenamefont {Chen},
  \citenamefont {Parkin}, \citenamefont {Wilson}, \citenamefont {Toberer},
  \citenamefont {McQueen},\ and\ \citenamefont
  {Ali}}]{doi:10.1126/sciadv.abb6003}%
  \BibitemOpen
  \bibfield  {author} {\bibinfo {author} {\bibfnamefont {S.-Y.}\ \bibnamefont
  {Yang}}, \bibinfo {author} {\bibfnamefont {Y.}~\bibnamefont {Wang}}, \bibinfo
  {author} {\bibfnamefont {B.~R.}\ \bibnamefont {Ortiz}}, \bibinfo {author}
  {\bibfnamefont {D.}~\bibnamefont {Liu}}, \bibinfo {author} {\bibfnamefont
  {J.}~\bibnamefont {Gayles}}, \bibinfo {author} {\bibfnamefont
  {E.}~\bibnamefont {Derunova}}, \bibinfo {author} {\bibfnamefont
  {R.}~\bibnamefont {Gonzalez-Hernandez}}, \bibinfo {author} {\bibfnamefont
  {L.}~\bibnamefont {Smejkal}}, \bibinfo {author} {\bibfnamefont
  {Y.}~\bibnamefont {Chen}}, \bibinfo {author} {\bibfnamefont {S.~S.~P.}\
  \bibnamefont {Parkin}}, \bibinfo {author} {\bibfnamefont {S.~D.}\
  \bibnamefont {Wilson}}, \bibinfo {author} {\bibfnamefont {E.~S.}\
  \bibnamefont {Toberer}}, \bibinfo {author} {\bibfnamefont {T.}~\bibnamefont
  {McQueen}},\ and\ \bibinfo {author} {\bibfnamefont {M.~N.}\ \bibnamefont
  {Ali}},\ }\bibfield  {title} {\bibinfo {title} {Giant, unconventional
  anomalous hall effect in the metallic frustrated magnet candidate,
  {${\mathrm{KV}}_{3}{\mathrm{Sb}}_{5}$}},\ }\href@noop {} {\bibfield
  {journal} {\bibinfo  {journal} {Sci. Adv.}\ }\textbf {\bibinfo {volume}
  {6}},\ \bibinfo {pages} {eabb6003} (\bibinfo {year} {2020})}\BibitemShut
  {NoStop}%
\bibitem [{\citenamefont {Ortiz}\ \emph {et~al.}(2019)\citenamefont {Ortiz},
  \citenamefont {Gomes}, \citenamefont {Morey}, \citenamefont {Winiarski},
  \citenamefont {Bordelon}, \citenamefont {Mangum}, \citenamefont {Oswald},
  \citenamefont {Rodriguez-Rivera}, \citenamefont {Neilson}, \citenamefont
  {Wilson}, \citenamefont {Ertekin}, \citenamefont {McQueen},\ and\
  \citenamefont {Toberer}}]{r5}%
  \BibitemOpen
  \bibfield  {author} {\bibinfo {author} {\bibfnamefont {B.~R.}\ \bibnamefont
  {Ortiz}}, \bibinfo {author} {\bibfnamefont {L.~C.}\ \bibnamefont {Gomes}},
  \bibinfo {author} {\bibfnamefont {J.~R.}\ \bibnamefont {Morey}}, \bibinfo
  {author} {\bibfnamefont {M.}~\bibnamefont {Winiarski}}, \bibinfo {author}
  {\bibfnamefont {M.}~\bibnamefont {Bordelon}}, \bibinfo {author}
  {\bibfnamefont {J.~S.}\ \bibnamefont {Mangum}}, \bibinfo {author}
  {\bibfnamefont {I.~W.~H.}\ \bibnamefont {Oswald}}, \bibinfo {author}
  {\bibfnamefont {J.~A.}\ \bibnamefont {Rodriguez-Rivera}}, \bibinfo {author}
  {\bibfnamefont {J.~R.}\ \bibnamefont {Neilson}}, \bibinfo {author}
  {\bibfnamefont {S.~D.}\ \bibnamefont {Wilson}}, \bibinfo {author}
  {\bibfnamefont {E.}~\bibnamefont {Ertekin}}, \bibinfo {author} {\bibfnamefont
  {T.~M.}\ \bibnamefont {McQueen}},\ and\ \bibinfo {author} {\bibfnamefont
  {E.~S.}\ \bibnamefont {Toberer}},\ }\bibfield  {title} {\bibinfo {title} {New
  kagome prototype materials: discovery of
  {${\mathrm{KV}}_{3}{\mathrm{Sb}}_{5},{\mathrm{RbV}}_{3}{\mathrm{Sb}}_{5}$,
  and ${\mathrm{CsV}}_{3}{\mathrm{Sb}}_{5}$}},\ }\href
  {https://doi.org/10.1103/PhysRevMaterials.3.094407} {\bibfield  {journal}
  {\bibinfo  {journal} {Phys. Rev. Mater.}\ }\textbf {\bibinfo {volume} {3}},\
  \bibinfo {pages} {094407} (\bibinfo {year} {2019})}\BibitemShut {NoStop}%
\bibitem [{\citenamefont {Kheirkhah}\ \emph {et~al.}(2022)\citenamefont
  {Kheirkhah}, \citenamefont {Zhu}, \citenamefont {Maciejko},\ and\
  \citenamefont {Yan}}]{PhysRevB.106.085420}%
  \BibitemOpen
  \bibfield  {author} {\bibinfo {author} {\bibfnamefont {M.}~\bibnamefont
  {Kheirkhah}}, \bibinfo {author} {\bibfnamefont {D.}~\bibnamefont {Zhu}},
  \bibinfo {author} {\bibfnamefont {J.}~\bibnamefont {Maciejko}},\ and\
  \bibinfo {author} {\bibfnamefont {Z.}~\bibnamefont {Yan}},\ }\bibfield
  {title} {\bibinfo {title} {Corner- and sublattice-sensitive majorana zero
  modes on the kagome lattice},\ }\href
  {https://doi.org/10.1103/PhysRevB.106.085420} {\bibfield  {journal} {\bibinfo
   {journal} {Phys. Rev. B}\ }\textbf {\bibinfo {volume} {106}},\ \bibinfo
  {pages} {085420} (\bibinfo {year} {2022})}\BibitemShut {NoStop}%
\bibitem [{\citenamefont {Han}\ \emph {et~al.}(2023)\citenamefont {Han},
  \citenamefont {Che}, \citenamefont {Ye},\ and\ \citenamefont {Huang}}]{r31}%
  \BibitemOpen
  \bibfield  {author} {\bibinfo {author} {\bibfnamefont {T.}~\bibnamefont
  {Han}}, \bibinfo {author} {\bibfnamefont {J.}~\bibnamefont {Che}}, \bibinfo
  {author} {\bibfnamefont {C.}~\bibnamefont {Ye}},\ and\ \bibinfo {author}
  {\bibfnamefont {H.}~\bibnamefont {Huang}},\ }\bibfield  {title} {\bibinfo
  {title} {Ginzburg-landau analysis on the physical properties of the kagome
  superconductor {${\mathrm{CsV}}_{3}{\mathrm{Sb}}_{5}$}},\ }\href
  {https://www.mdpi.com/2073-4352/13/2/321} {\bibfield  {journal} {\bibinfo
  {journal} {Crystals}\ }\textbf {\bibinfo {volume} {13}} (\bibinfo {year}
  {2023})}\BibitemShut {NoStop}%
\bibitem [{\citenamefont {Wang}\ \emph {et~al.}(2013)\citenamefont {Wang},
  \citenamefont {Li}, \citenamefont {Xiang},\ and\ \citenamefont {Wang}}]{r4}%
  \BibitemOpen
  \bibfield  {author} {\bibinfo {author} {\bibfnamefont {W.-S.}\ \bibnamefont
  {Wang}}, \bibinfo {author} {\bibfnamefont {Z.-Z.}\ \bibnamefont {Li}},
  \bibinfo {author} {\bibfnamefont {Y.-Y.}\ \bibnamefont {Xiang}},\ and\
  \bibinfo {author} {\bibfnamefont {Q.-H.}\ \bibnamefont {Wang}},\ }\bibfield
  {title} {\bibinfo {title} {Competing electronic orders on kagome lattices at
  van hove filling},\ }\href {https://doi.org/10.1103/PhysRevB.87.115135}
  {\bibfield  {journal} {\bibinfo  {journal} {Phys. Rev. B}\ }\textbf {\bibinfo
  {volume} {87}},\ \bibinfo {pages} {115135} (\bibinfo {year}
  {2013})}\BibitemShut {NoStop}%
\bibitem [{\citenamefont {Tazai}\ \emph {et~al.}(2022)\citenamefont {Tazai},
  \citenamefont {Yamakawa}, \citenamefont {Onari},\ and\ \citenamefont
  {Kontani}}]{r30}%
  \BibitemOpen
  \bibfield  {author} {\bibinfo {author} {\bibfnamefont {R.}~\bibnamefont
  {Tazai}}, \bibinfo {author} {\bibfnamefont {Y.}~\bibnamefont {Yamakawa}},
  \bibinfo {author} {\bibfnamefont {S.}~\bibnamefont {Onari}},\ and\ \bibinfo
  {author} {\bibfnamefont {H.}~\bibnamefont {Kontani}},\ }\bibfield  {title}
  {\bibinfo {title} {Mechanism of exotic density-wave and beyond-migdal
  unconventional superconductivity in kagome metal
  {$A{\mathrm{V}}_{3}{\mathrm{Sb}}_{5}$
  ($A=\mathrm{K},\mathrm{Rb},\mathrm{Cs}$)}},\ }\href
  {https://www.science.org/doi/abs/10.1126/sciadv.abl4108} {\bibfield
  {journal} {\bibinfo  {journal} {Sci. Adv.}\ }\textbf {\bibinfo {volume}
  {8}},\ \bibinfo {pages} {eabl4108} (\bibinfo {year} {2022})}\BibitemShut
  {NoStop}%
\bibitem [{\citenamefont {Jiang}\ \emph {et~al.}(2022)\citenamefont {Jiang},
  \citenamefont {Yu},\ and\ \citenamefont {Pan}}]{PhysRevB.106.014501}%
  \BibitemOpen
  \bibfield  {author} {\bibinfo {author} {\bibfnamefont {H.-M.}\ \bibnamefont
  {Jiang}}, \bibinfo {author} {\bibfnamefont {S.-L.}\ \bibnamefont {Yu}},\ and\
  \bibinfo {author} {\bibfnamefont {X.-Y.}\ \bibnamefont {Pan}},\ }\bibfield
  {title} {\bibinfo {title} {Electronic structure and spin-lattice relaxation
  in superconducting vortex states on the kagome lattice near van hove
  filling},\ }\href {https://doi.org/10.1103/PhysRevB.106.014501} {\bibfield
  {journal} {\bibinfo  {journal} {Phys. Rev. B}\ }\textbf {\bibinfo {volume}
  {106}},\ \bibinfo {pages} {014501} (\bibinfo {year} {2022})}\BibitemShut
  {NoStop}%
\bibitem [{\citenamefont {Yu}\ and\ \citenamefont {Li}(2012)}]{r2}%
  \BibitemOpen
  \bibfield  {author} {\bibinfo {author} {\bibfnamefont {S.-L.}\ \bibnamefont
  {Yu}}\ and\ \bibinfo {author} {\bibfnamefont {J.-X.}\ \bibnamefont {Li}},\
  }\bibfield  {title} {\bibinfo {title} {Chiral superconducting phase and
  chiral spin-density-wave phase in a hubbard model on the kagome lattice},\
  }\href {https://doi.org/10.1103/PhysRevB.85.144402} {\bibfield  {journal}
  {\bibinfo  {journal} {Phys. Rev. B}\ }\textbf {\bibinfo {volume} {85}},\
  \bibinfo {pages} {144402} (\bibinfo {year} {2012})}\BibitemShut {NoStop}%
\bibitem [{\citenamefont {Kiesel}\ and\ \citenamefont {Thomale}(2012)}]{r27}%
  \BibitemOpen
  \bibfield  {author} {\bibinfo {author} {\bibfnamefont {M.~L.}\ \bibnamefont
  {Kiesel}}\ and\ \bibinfo {author} {\bibfnamefont {R.}~\bibnamefont
  {Thomale}},\ }\bibfield  {title} {\bibinfo {title} {Sublattice interference
  in the kagome hubbard model},\ }\href
  {https://doi.org/10.1103/PhysRevB.86.121105} {\bibfield  {journal} {\bibinfo
  {journal} {Phys. Rev. B}\ }\textbf {\bibinfo {volume} {86}},\ \bibinfo
  {pages} {121105} (\bibinfo {year} {2012})}\BibitemShut {NoStop}%
\bibitem [{\citenamefont {Wen}\ \emph {et~al.}(2022)\citenamefont {Wen},
  \citenamefont {Zhu}, \citenamefont {Xiao}, \citenamefont {Hao}, \citenamefont
  {Mondaini}, \citenamefont {Guo},\ and\ \citenamefont {Feng}}]{r28}%
  \BibitemOpen
  \bibfield  {author} {\bibinfo {author} {\bibfnamefont {C.}~\bibnamefont
  {Wen}}, \bibinfo {author} {\bibfnamefont {X.}~\bibnamefont {Zhu}}, \bibinfo
  {author} {\bibfnamefont {Z.}~\bibnamefont {Xiao}}, \bibinfo {author}
  {\bibfnamefont {N.}~\bibnamefont {Hao}}, \bibinfo {author} {\bibfnamefont
  {R.}~\bibnamefont {Mondaini}}, \bibinfo {author} {\bibfnamefont
  {H.}~\bibnamefont {Guo}},\ and\ \bibinfo {author} {\bibfnamefont
  {S.}~\bibnamefont {Feng}},\ }\bibfield  {title} {\bibinfo {title}
  {Superconducting pairing symmetry in the kagome-lattice hubbard model},\
  }\href {https://doi.org/10.1103/PhysRevB.105.075118} {\bibfield  {journal}
  {\bibinfo  {journal} {Phys. Rev. B}\ }\textbf {\bibinfo {volume} {105}},\
  \bibinfo {pages} {075118} (\bibinfo {year} {2022})}\BibitemShut {NoStop}%
\bibitem [{\citenamefont {Wu}\ \emph {et~al.}(2021)\citenamefont {Wu},
  \citenamefont {Schwemmer}, \citenamefont {M\"uller}, \citenamefont
  {Consiglio}, \citenamefont {Sangiovanni}, \citenamefont {Di~Sante},
  \citenamefont {Iqbal}, \citenamefont {Hanke}, \citenamefont {Schnyder},
  \citenamefont {Denner}, \citenamefont {Fischer}, \citenamefont {Neupert},\
  and\ \citenamefont {Thomale}}]{r29}%
  \BibitemOpen
  \bibfield  {author} {\bibinfo {author} {\bibfnamefont {X.}~\bibnamefont
  {Wu}}, \bibinfo {author} {\bibfnamefont {T.}~\bibnamefont {Schwemmer}},
  \bibinfo {author} {\bibfnamefont {T.}~\bibnamefont {M\"uller}}, \bibinfo
  {author} {\bibfnamefont {A.}~\bibnamefont {Consiglio}}, \bibinfo {author}
  {\bibfnamefont {G.}~\bibnamefont {Sangiovanni}}, \bibinfo {author}
  {\bibfnamefont {D.}~\bibnamefont {Di~Sante}}, \bibinfo {author}
  {\bibfnamefont {Y.}~\bibnamefont {Iqbal}}, \bibinfo {author} {\bibfnamefont
  {W.}~\bibnamefont {Hanke}}, \bibinfo {author} {\bibfnamefont {A.~P.}\
  \bibnamefont {Schnyder}}, \bibinfo {author} {\bibfnamefont {M.~M.}\
  \bibnamefont {Denner}}, \bibinfo {author} {\bibfnamefont {M.~H.}\
  \bibnamefont {Fischer}}, \bibinfo {author} {\bibfnamefont {T.}~\bibnamefont
  {Neupert}},\ and\ \bibinfo {author} {\bibfnamefont {R.}~\bibnamefont
  {Thomale}},\ }\bibfield  {title} {\bibinfo {title} {Nature of unconventional
  pairing in the kagome superconductors {$A{\mathrm{V}}_{3}{\mathrm{Sb}}_{5}$
  ($A=\mathrm{K},\mathrm{Rb},\mathrm{Cs}$)}},\ }\href
  {https://doi.org/10.1103/PhysRevLett.127.177001} {\bibfield  {journal}
  {\bibinfo  {journal} {Phys. Rev. Lett.}\ }\textbf {\bibinfo {volume} {127}},\
  \bibinfo {pages} {177001} (\bibinfo {year} {2021})}\BibitemShut {NoStop}%
\bibitem [{\citenamefont {Kiesel}\ \emph {et~al.}(2013)\citenamefont {Kiesel},
  \citenamefont {Platt},\ and\ \citenamefont {Thomale}}]{r3}%
  \BibitemOpen
  \bibfield  {author} {\bibinfo {author} {\bibfnamefont {M.~L.}\ \bibnamefont
  {Kiesel}}, \bibinfo {author} {\bibfnamefont {C.}~\bibnamefont {Platt}},\ and\
  \bibinfo {author} {\bibfnamefont {R.}~\bibnamefont {Thomale}},\ }\bibfield
  {title} {\bibinfo {title} {Unconventional fermi surface instabilities in the
  kagome hubbard model},\ }\href
  {https://doi.org/10.1103/PhysRevLett.110.126405} {\bibfield  {journal}
  {\bibinfo  {journal} {Phys. Rev. Lett.}\ }\textbf {\bibinfo {volume} {110}},\
  \bibinfo {pages} {126405} (\bibinfo {year} {2013})}\BibitemShut {NoStop}%
\bibitem [{\citenamefont {Ding}\ \emph {et~al.}(2022)\citenamefont {Ding},
  \citenamefont {Lee}, \citenamefont {Wu},\ and\ \citenamefont
  {Thomale}}]{PhysRevB.105.174518}%
  \BibitemOpen
  \bibfield  {author} {\bibinfo {author} {\bibfnamefont {P.}~\bibnamefont
  {Ding}}, \bibinfo {author} {\bibfnamefont {C.~H.}\ \bibnamefont {Lee}},
  \bibinfo {author} {\bibfnamefont {X.}~\bibnamefont {Wu}},\ and\ \bibinfo
  {author} {\bibfnamefont {R.}~\bibnamefont {Thomale}},\ }\bibfield  {title}
  {\bibinfo {title} {Diagnosis of pairing symmetry by vortex and edge spectra
  in kagome superconductors},\ }\href
  {https://doi.org/10.1103/PhysRevB.105.174518} {\bibfield  {journal} {\bibinfo
   {journal} {Phys. Rev. B}\ }\textbf {\bibinfo {volume} {105}},\ \bibinfo
  {pages} {174518} (\bibinfo {year} {2022})}\BibitemShut {NoStop}%
\bibitem [{\citenamefont {Mu}\ \emph {et~al.}(2021)\citenamefont {Mu},
  \citenamefont {Yin}, \citenamefont {Tu}, \citenamefont {Gong}, \citenamefont
  {Lei}, \citenamefont {Li},\ and\ \citenamefont {Luo}}]{r19}%
  \BibitemOpen
  \bibfield  {author} {\bibinfo {author} {\bibfnamefont {C.}~\bibnamefont
  {Mu}}, \bibinfo {author} {\bibfnamefont {Q.}~\bibnamefont {Yin}}, \bibinfo
  {author} {\bibfnamefont {Z.}~\bibnamefont {Tu}}, \bibinfo {author}
  {\bibfnamefont {C.}~\bibnamefont {Gong}}, \bibinfo {author} {\bibfnamefont
  {H.}~\bibnamefont {Lei}}, \bibinfo {author} {\bibfnamefont {Z.}~\bibnamefont
  {Li}},\ and\ \bibinfo {author} {\bibfnamefont {J.}~\bibnamefont {Luo}},\
  }\bibfield  {title} {\bibinfo {title} {S-wave superconductivity in kagome
  metal {${\mathrm{CsV}}_{3}{\mathrm{Sb}}_{5}$} revealed by {$^{121/123}$Sb NQR
  and $^{51}$V NMR} measurements},\ }\href
  {https://doi.org/10.1088/0256-307X/38/7/077402} {\bibfield  {journal}
  {\bibinfo  {journal} {Chin. Phys. Lett.}\ }\textbf {\bibinfo {volume} {38}},\
  \bibinfo {eid} {077402} (\bibinfo {year} {2021})}\BibitemShut {NoStop}%
\bibitem [{\citenamefont {Zhang}\ \emph {et~al.}(2023)\citenamefont {Zhang},
  \citenamefont {Liu}, \citenamefont {Wang}, \citenamefont {Tsang},
  \citenamefont {Wang}, \citenamefont {Lam}, \citenamefont {Wang},
  \citenamefont {Xie}, \citenamefont {Zhou}, \citenamefont {Zhao},
  \citenamefont {Wang}, \citenamefont {Tallon}, \citenamefont {Lai},\ and\
  \citenamefont {Goh}}]{r20}%
  \BibitemOpen
  \bibfield  {author} {\bibinfo {author} {\bibfnamefont {W.}~\bibnamefont
  {Zhang}}, \bibinfo {author} {\bibfnamefont {X.}~\bibnamefont {Liu}}, \bibinfo
  {author} {\bibfnamefont {L.}~\bibnamefont {Wang}}, \bibinfo {author}
  {\bibfnamefont {C.~W.}\ \bibnamefont {Tsang}}, \bibinfo {author}
  {\bibfnamefont {Z.}~\bibnamefont {Wang}}, \bibinfo {author} {\bibfnamefont
  {S.~T.}\ \bibnamefont {Lam}}, \bibinfo {author} {\bibfnamefont
  {W.}~\bibnamefont {Wang}}, \bibinfo {author} {\bibfnamefont {J.}~\bibnamefont
  {Xie}}, \bibinfo {author} {\bibfnamefont {X.}~\bibnamefont {Zhou}}, \bibinfo
  {author} {\bibfnamefont {Y.}~\bibnamefont {Zhao}}, \bibinfo {author}
  {\bibfnamefont {S.}~\bibnamefont {Wang}}, \bibinfo {author} {\bibfnamefont
  {J.}~\bibnamefont {Tallon}}, \bibinfo {author} {\bibfnamefont {K.~T.}\
  \bibnamefont {Lai}},\ and\ \bibinfo {author} {\bibfnamefont {S.~K.}\
  \bibnamefont {Goh}},\ }\bibfield  {title} {\bibinfo {title} {Nodeless
  superconductivity in kagome metal {${\mathrm{CsV}}_{3}{\mathrm{Sb}}_{5}$}
  with and without time reversal symmetry breaking},\ }\href
  {https://doi.org/10.1021/acs.nanolett.2c04103} {\bibfield  {journal}
  {\bibinfo  {journal} {Nano Lett.}\ }\textbf {\bibinfo {volume} {23}},\
  \bibinfo {pages} {872} (\bibinfo {year} {2023})}\BibitemShut {NoStop}%
\bibitem [{\citenamefont {Duan}\ \emph {et~al.}(2021)\citenamefont {Duan},
  \citenamefont {Nie}, \citenamefont {Luo}, \citenamefont {Yu}, \citenamefont
  {R.Ortiz}, \citenamefont {Yin}, \citenamefont {Su}, \citenamefont {Du},
  \citenamefont {Wang}, \citenamefont {Chen}, \citenamefont {Lu}, \citenamefont
  {Ying}, \citenamefont {D.Wilson}, \citenamefont {Chen}, \citenamefont
  {Song},\ and\ \citenamefont {Yuan}}]{r24}%
  \BibitemOpen
  \bibfield  {author} {\bibinfo {author} {\bibfnamefont {W.}~\bibnamefont
  {Duan}}, \bibinfo {author} {\bibfnamefont {Z.}~\bibnamefont {Nie}}, \bibinfo
  {author} {\bibfnamefont {S.}~\bibnamefont {Luo}}, \bibinfo {author}
  {\bibfnamefont {F.}~\bibnamefont {Yu}}, \bibinfo {author} {\bibfnamefont
  {B.}~\bibnamefont {R.Ortiz}}, \bibinfo {author} {\bibfnamefont
  {L.}~\bibnamefont {Yin}}, \bibinfo {author} {\bibfnamefont {H.}~\bibnamefont
  {Su}}, \bibinfo {author} {\bibfnamefont {F.}~\bibnamefont {Du}}, \bibinfo
  {author} {\bibfnamefont {A.}~\bibnamefont {Wang}}, \bibinfo {author}
  {\bibfnamefont {Y.}~\bibnamefont {Chen}}, \bibinfo {author} {\bibfnamefont
  {X.}~\bibnamefont {Lu}}, \bibinfo {author} {\bibfnamefont {J.}~\bibnamefont
  {Ying}}, \bibinfo {author} {\bibfnamefont {S.}~\bibnamefont {D.Wilson}},
  \bibinfo {author} {\bibfnamefont {X.}~\bibnamefont {Chen}}, \bibinfo {author}
  {\bibfnamefont {Y.}~\bibnamefont {Song}},\ and\ \bibinfo {author}
  {\bibfnamefont {H.}~\bibnamefont {Yuan}},\ }\bibfield  {title} {\bibinfo
  {title} {Nodeless superconductivity in the kagome metal
  {${\mathrm{CsV}}_{3}{\mathrm{Sb}}_{5}$}},\ }\href@noop {} {\bibfield
  {journal} {\bibinfo  {journal} {Sci. China Phys. Mech. Astron.}\ }\textbf
  {\bibinfo {volume} {64}},\ \bibinfo {pages} {15} (\bibinfo {year}
  {2021})}\BibitemShut {NoStop}%
\bibitem [{\citenamefont {Ni}\ \emph {et~al.}(2022)\citenamefont {Ni},
  \citenamefont {Ma}, \citenamefont {Zhang}, \citenamefont {Yuan},
  \citenamefont {Yang}, \citenamefont {Lu}, \citenamefont {Wang}, \citenamefont
  {Sun}, \citenamefont {Zhao}, \citenamefont {Li}, \citenamefont {Liu},
  \citenamefont {Zhang}, \citenamefont {Chen}, \citenamefont {Jin},
  \citenamefont {Cheng}, \citenamefont {Yu}, \citenamefont {Zhou},
  \citenamefont {Dong}, \citenamefont {Hu}, \citenamefont {Gao},\ and\
  \citenamefont {Zhao}}]{r25}%
  \BibitemOpen
  \bibfield  {author} {\bibinfo {author} {\bibfnamefont {S.}~\bibnamefont
  {Ni}}, \bibinfo {author} {\bibfnamefont {S.}~\bibnamefont {Ma}}, \bibinfo
  {author} {\bibfnamefont {Y.}~\bibnamefont {Zhang}}, \bibinfo {author}
  {\bibfnamefont {J.}~\bibnamefont {Yuan}}, \bibinfo {author} {\bibfnamefont
  {H.}~\bibnamefont {Yang}}, \bibinfo {author} {\bibfnamefont {Z.}~\bibnamefont
  {Lu}}, \bibinfo {author} {\bibfnamefont {N.}~\bibnamefont {Wang}}, \bibinfo
  {author} {\bibfnamefont {J.}~\bibnamefont {Sun}}, \bibinfo {author}
  {\bibfnamefont {Z.}~\bibnamefont {Zhao}}, \bibinfo {author} {\bibfnamefont
  {D.}~\bibnamefont {Li}}, \bibinfo {author} {\bibfnamefont {S.}~\bibnamefont
  {Liu}}, \bibinfo {author} {\bibfnamefont {H.}~\bibnamefont {Zhang}}, \bibinfo
  {author} {\bibfnamefont {H.}~\bibnamefont {Chen}}, \bibinfo {author}
  {\bibfnamefont {K.}~\bibnamefont {Jin}}, \bibinfo {author} {\bibfnamefont
  {J.}~\bibnamefont {Cheng}}, \bibinfo {author} {\bibfnamefont
  {L.}~\bibnamefont {Yu}}, \bibinfo {author} {\bibfnamefont {F.}~\bibnamefont
  {Zhou}}, \bibinfo {author} {\bibfnamefont {X.}~\bibnamefont {Dong}}, \bibinfo
  {author} {\bibfnamefont {J.}~\bibnamefont {Hu}}, \bibinfo {author}
  {\bibfnamefont {H.-J.}\ \bibnamefont {Gao}},\ and\ \bibinfo {author}
  {\bibfnamefont {Z.}~\bibnamefont {Zhao}},\ }\bibfield  {title} {\bibinfo
  {title} {Microscopic evidence for anisotropic multigap superconductivity in
  the {${\mathrm{CsV}}_{3}{\mathrm{Sb}}_{5}$} kagome superconductor},\
  }\href@noop {} {\bibfield  {journal} {\bibinfo  {journal} {npj Quantum
  Mater.}\ }\textbf {\bibinfo {volume} {7}} (\bibinfo {year}
  {2022})}\BibitemShut {NoStop}%
\bibitem [{\citenamefont {Roppongi}\ \emph {et~al.}(2023)\citenamefont
  {Roppongi}, \citenamefont {Ishihara}, \citenamefont {Tanaka}, \citenamefont
  {Ogawa}, \citenamefont {Okada}, \citenamefont {Liu}, \citenamefont {Mukasa},
  \citenamefont {Mizukami}, \citenamefont {Uwatoko}, \citenamefont {Grasset},
  \citenamefont {Konczykowski}, \citenamefont {Ortiz}, \citenamefont {Wilson},
  \citenamefont {Hashimoto},\ and\ \citenamefont {Shibauchi}}]{r26}%
  \BibitemOpen
  \bibfield  {author} {\bibinfo {author} {\bibfnamefont {M.}~\bibnamefont
  {Roppongi}}, \bibinfo {author} {\bibfnamefont {K.}~\bibnamefont {Ishihara}},
  \bibinfo {author} {\bibfnamefont {Y.}~\bibnamefont {Tanaka}}, \bibinfo
  {author} {\bibfnamefont {K.}~\bibnamefont {Ogawa}}, \bibinfo {author}
  {\bibfnamefont {K.}~\bibnamefont {Okada}}, \bibinfo {author} {\bibfnamefont
  {S.}~\bibnamefont {Liu}}, \bibinfo {author} {\bibfnamefont {K.}~\bibnamefont
  {Mukasa}}, \bibinfo {author} {\bibfnamefont {Y.}~\bibnamefont {Mizukami}},
  \bibinfo {author} {\bibfnamefont {Y.}~\bibnamefont {Uwatoko}}, \bibinfo
  {author} {\bibfnamefont {R.}~\bibnamefont {Grasset}}, \bibinfo {author}
  {\bibfnamefont {M.}~\bibnamefont {Konczykowski}}, \bibinfo {author}
  {\bibfnamefont {B.~R.}\ \bibnamefont {Ortiz}}, \bibinfo {author}
  {\bibfnamefont {S.~D.}\ \bibnamefont {Wilson}}, \bibinfo {author}
  {\bibfnamefont {K.}~\bibnamefont {Hashimoto}},\ and\ \bibinfo {author}
  {\bibfnamefont {T.}~\bibnamefont {Shibauchi}},\ }\bibfield  {title} {\bibinfo
  {title} {Bulk evidence of anisotropic s-wave pairing with no sign change in
  the kagome superconductor {${\mathrm{CsV}}_{3}{\mathrm{Sb}}_{5}$}},\
  }\href@noop {} {\bibfield  {journal} {\bibinfo  {journal} {Nat. Commun.}\
  }\textbf {\bibinfo {volume} {14}},\ \bibinfo {pages} {667} (\bibinfo {year}
  {2023})}\BibitemShut {NoStop}%
\bibitem [{\citenamefont {Xu}\ \emph {et~al.}(2021)\citenamefont {Xu},
  \citenamefont {Yan}, \citenamefont {Yin}, \citenamefont {Xia}, \citenamefont
  {Fang}, \citenamefont {Chen}, \citenamefont {Li}, \citenamefont {Yang},
  \citenamefont {Guo},\ and\ \citenamefont {Feng}}]{r23}%
  \BibitemOpen
  \bibfield  {author} {\bibinfo {author} {\bibfnamefont {H.-S.}\ \bibnamefont
  {Xu}}, \bibinfo {author} {\bibfnamefont {Y.-J.}\ \bibnamefont {Yan}},
  \bibinfo {author} {\bibfnamefont {R.}~\bibnamefont {Yin}}, \bibinfo {author}
  {\bibfnamefont {W.}~\bibnamefont {Xia}}, \bibinfo {author} {\bibfnamefont
  {S.}~\bibnamefont {Fang}}, \bibinfo {author} {\bibfnamefont {Z.}~\bibnamefont
  {Chen}}, \bibinfo {author} {\bibfnamefont {Y.}~\bibnamefont {Li}}, \bibinfo
  {author} {\bibfnamefont {W.}~\bibnamefont {Yang}}, \bibinfo {author}
  {\bibfnamefont {Y.}~\bibnamefont {Guo}},\ and\ \bibinfo {author}
  {\bibfnamefont {D.-L.}\ \bibnamefont {Feng}},\ }\bibfield  {title} {\bibinfo
  {title} {Multiband superconductivity with sign-preserving order parameter in
  kagome superconductor {${\mathrm{CsV}}_{3}{\mathrm{Sb}}_{5}$}},\ }\href
  {https://doi.org/10.1103/PhysRevLett.127.187004} {\bibfield  {journal}
  {\bibinfo  {journal} {Phys. Rev. Lett.}\ }\textbf {\bibinfo {volume} {127}},\
  \bibinfo {pages} {187004} (\bibinfo {year} {2021})}\BibitemShut {NoStop}%
\bibitem [{\citenamefont {Zhong}\ \emph {et~al.}(2023)\citenamefont {Zhong},
  \citenamefont {Liu}, \citenamefont {Wu}, \citenamefont {Guguchia},
  \citenamefont {Yin}, \citenamefont {Mine}, \citenamefont {Li}, \citenamefont
  {Najafzadeh}, \citenamefont {Das}, \citenamefont {Mielke}, \citenamefont
  {Khasanov}, \citenamefont {Luetkens}, \citenamefont {Suzuki}, \citenamefont
  {Liu}, \citenamefont {Han}, \citenamefont {Kondo}, \citenamefont {Hu},
  \citenamefont {Shin}, \citenamefont {Wang}, \citenamefont {Shi},
  \citenamefont {Yao},\ and\ \citenamefont {Okazaki}}]{Zhong2023}%
  \BibitemOpen
  \bibfield  {author} {\bibinfo {author} {\bibfnamefont {Y.}~\bibnamefont
  {Zhong}}, \bibinfo {author} {\bibfnamefont {J.}~\bibnamefont {Liu}}, \bibinfo
  {author} {\bibfnamefont {X.}~\bibnamefont {Wu}}, \bibinfo {author}
  {\bibfnamefont {Z.}~\bibnamefont {Guguchia}}, \bibinfo {author}
  {\bibfnamefont {J.-X.}\ \bibnamefont {Yin}}, \bibinfo {author} {\bibfnamefont
  {A.}~\bibnamefont {Mine}}, \bibinfo {author} {\bibfnamefont {Y.}~\bibnamefont
  {Li}}, \bibinfo {author} {\bibfnamefont {S.}~\bibnamefont {Najafzadeh}},
  \bibinfo {author} {\bibfnamefont {D.}~\bibnamefont {Das}}, \bibinfo {author}
  {\bibfnamefont {C.}~\bibnamefont {Mielke}}, \bibinfo {author} {\bibfnamefont
  {R.}~\bibnamefont {Khasanov}}, \bibinfo {author} {\bibfnamefont
  {H.}~\bibnamefont {Luetkens}}, \bibinfo {author} {\bibfnamefont
  {T.}~\bibnamefont {Suzuki}}, \bibinfo {author} {\bibfnamefont
  {K.}~\bibnamefont {Liu}}, \bibinfo {author} {\bibfnamefont {X.}~\bibnamefont
  {Han}}, \bibinfo {author} {\bibfnamefont {T.}~\bibnamefont {Kondo}}, \bibinfo
  {author} {\bibfnamefont {J.}~\bibnamefont {Hu}}, \bibinfo {author}
  {\bibfnamefont {S.}~\bibnamefont {Shin}}, \bibinfo {author} {\bibfnamefont
  {Z.}~\bibnamefont {Wang}}, \bibinfo {author} {\bibfnamefont {X.}~\bibnamefont
  {Shi}}, \bibinfo {author} {\bibfnamefont {Y.}~\bibnamefont {Yao}},\ and\
  \bibinfo {author} {\bibfnamefont {K.}~\bibnamefont {Okazaki}},\ }\bibfield
  {title} {\bibinfo {title} {Nodeless electron pairing in
  {${\mathrm{CsV}}_{3}{\mathrm{Sb}}_{5}$}-derived kagome superconductors},\
  }\href {https://doi.org/10.1038/s41586-023-05907-x} {\bibfield  {journal}
  {\bibinfo  {journal} {Nature}\ }\textbf {\bibinfo {volume} {617}},\ \bibinfo
  {pages} {488} (\bibinfo {year} {2023})}\BibitemShut {NoStop}%
\bibitem [{\citenamefont {Balatsky}\ \emph {et~al.}(2006)\citenamefont
  {Balatsky}, \citenamefont {Vekhter},\ and\ \citenamefont
  {Zhu}}]{RevModPhys.78.373}%
  \BibitemOpen
  \bibfield  {author} {\bibinfo {author} {\bibfnamefont {A.~V.}\ \bibnamefont
  {Balatsky}}, \bibinfo {author} {\bibfnamefont {I.}~\bibnamefont {Vekhter}},\
  and\ \bibinfo {author} {\bibfnamefont {J.-X.}\ \bibnamefont {Zhu}},\
  }\bibfield  {title} {\bibinfo {title} {Impurity-induced states in
  conventional and unconventional superconductors},\ }\href
  {https://doi.org/10.1103/RevModPhys.78.373} {\bibfield  {journal} {\bibinfo
  {journal} {Rev. Mod. Phys.}\ }\textbf {\bibinfo {volume} {78}},\ \bibinfo
  {pages} {373} (\bibinfo {year} {2006})}\BibitemShut {NoStop}%
\bibitem [{\citenamefont {Hu}(1994)}]{PhysRevLett.72.1526}%
  \BibitemOpen
  \bibfield  {author} {\bibinfo {author} {\bibfnamefont {C.-R.}\ \bibnamefont
  {Hu}},\ }\bibfield  {title} {\bibinfo {title} {Midgap surface states as a
  novel signature for
  ${\mathit{d}}_{\mathit{x}\mathit{a}}^{2}$-${\mathit{x}}_{\mathit{b}}^{2}$-wave
  superconductivity},\ }\href {https://doi.org/10.1103/PhysRevLett.72.1526}
  {\bibfield  {journal} {\bibinfo  {journal} {Phys. Rev. Lett.}\ }\textbf
  {\bibinfo {volume} {72}},\ \bibinfo {pages} {1526} (\bibinfo {year}
  {1994})}\BibitemShut {NoStop}%
\bibitem [{\citenamefont {Zhang}(2009)}]{PhysRevLett.103.186402}%
  \BibitemOpen
  \bibfield  {author} {\bibinfo {author} {\bibfnamefont {D.}~\bibnamefont
  {Zhang}},\ }\bibfield  {title} {\bibinfo {title} {Nonmagnetic impurity
  resonances as a signature of sign-reversal pairing in feas-based
  superconductors},\ }\href {https://doi.org/10.1103/PhysRevLett.103.186402}
  {\bibfield  {journal} {\bibinfo  {journal} {Phys. Rev. Lett.}\ }\textbf
  {\bibinfo {volume} {103}},\ \bibinfo {pages} {186402} (\bibinfo {year}
  {2009})}\BibitemShut {NoStop}%
\bibitem [{\citenamefont {Tsai}\ \emph {et~al.}(2009)\citenamefont {Tsai},
  \citenamefont {Zhang}, \citenamefont {Fang},\ and\ \citenamefont
  {Hu}}]{PhysRevB.80.064513}%
  \BibitemOpen
  \bibfield  {author} {\bibinfo {author} {\bibfnamefont {W.-F.}\ \bibnamefont
  {Tsai}}, \bibinfo {author} {\bibfnamefont {Y.-Y.}\ \bibnamefont {Zhang}},
  \bibinfo {author} {\bibfnamefont {C.}~\bibnamefont {Fang}},\ and\ \bibinfo
  {author} {\bibfnamefont {J.}~\bibnamefont {Hu}},\ }\bibfield  {title}
  {\bibinfo {title} {Impurity-induced bound states in iron-based
  superconductors with $s$-wave $\text{cos}\text{
  }{k}_{x}\ensuremath{\cdot}\text{cos}\text{ }{k}_{y}$ pairing symmetry},\
  }\href {https://doi.org/10.1103/PhysRevB.80.064513} {\bibfield  {journal}
  {\bibinfo  {journal} {Phys. Rev. B}\ }\textbf {\bibinfo {volume} {80}},\
  \bibinfo {pages} {064513} (\bibinfo {year} {2009})}\BibitemShut {NoStop}%
\bibitem [{\citenamefont {Li}\ and\ \citenamefont {Zhou}(2021)}]{Li2021}%
  \BibitemOpen
  \bibfield  {author} {\bibinfo {author} {\bibfnamefont {Y.-Q.}\ \bibnamefont
  {Li}}\ and\ \bibinfo {author} {\bibfnamefont {T.}~\bibnamefont {Zhou}},\
  }\bibfield  {title} {\bibinfo {title} {Impurity effect as a probe for the
  pairing symmetry of graphene-based superconductors},\ }\href
  {https://doi.org/10.1007/s11467-021-1056-y} {\bibfield  {journal} {\bibinfo
  {journal} {Front. Phys.}\ }\textbf {\bibinfo {volume} {16}},\ \bibinfo
  {pages} {43502} (\bibinfo {year} {2021})}\BibitemShut {NoStop}%
\bibitem [{\citenamefont {Xu}\ \emph {et~al.}(1995)\citenamefont {Xu},
  \citenamefont {Yip},\ and\ \citenamefont {Sauls}}]{PhysRevB.51.16233}%
  \BibitemOpen
  \bibfield  {author} {\bibinfo {author} {\bibfnamefont {D.}~\bibnamefont
  {Xu}}, \bibinfo {author} {\bibfnamefont {S.~K.}\ \bibnamefont {Yip}},\ and\
  \bibinfo {author} {\bibfnamefont {J.~A.}\ \bibnamefont {Sauls}},\ }\bibfield
  {title} {\bibinfo {title} {Nonlinear meissner effect in unconventional
  superconductors},\ }\href {https://doi.org/10.1103/PhysRevB.51.16233}
  {\bibfield  {journal} {\bibinfo  {journal} {Phys. Rev. B}\ }\textbf {\bibinfo
  {volume} {51}},\ \bibinfo {pages} {16233} (\bibinfo {year}
  {1995})}\BibitemShut {NoStop}%
\bibitem [{\citenamefont {Tsuei}\ and\ \citenamefont
  {Kirtley}(2000)}]{RevModPhys.72.969}%
  \BibitemOpen
  \bibfield  {author} {\bibinfo {author} {\bibfnamefont {C.~C.}\ \bibnamefont
  {Tsuei}}\ and\ \bibinfo {author} {\bibfnamefont {J.~R.}\ \bibnamefont
  {Kirtley}},\ }\bibfield  {title} {\bibinfo {title} {Pairing symmetry in
  cuprate superconductors},\ }\href {https://doi.org/10.1103/RevModPhys.72.969}
  {\bibfield  {journal} {\bibinfo  {journal} {Rev. Mod. Phys.}\ }\textbf
  {\bibinfo {volume} {72}},\ \bibinfo {pages} {969} (\bibinfo {year}
  {2000})}\BibitemShut {NoStop}%
\bibitem [{\citenamefont {Black-Schaffer}\ \emph {et~al.}(2014)\citenamefont
  {Black-Schaffer}, \citenamefont {Wu},\ and\ \citenamefont
  {Le~Hur}}]{PhysRevB.90.054521}%
  \BibitemOpen
  \bibfield  {author} {\bibinfo {author} {\bibfnamefont {A.~M.}\ \bibnamefont
  {Black-Schaffer}}, \bibinfo {author} {\bibfnamefont {W.}~\bibnamefont {Wu}},\
  and\ \bibinfo {author} {\bibfnamefont {K.}~\bibnamefont {Le~Hur}},\
  }\bibfield  {title} {\bibinfo {title} {Chiral $d$-wave superconductivity on
  the honeycomb lattice close to the mott state},\ }\href
  {https://doi.org/10.1103/PhysRevB.90.054521} {\bibfield  {journal} {\bibinfo
  {journal} {Phys. Rev. B}\ }\textbf {\bibinfo {volume} {90}},\ \bibinfo
  {pages} {054521} (\bibinfo {year} {2014})}\BibitemShut {NoStop}%
\bibitem [{\citenamefont {Faye}\ \emph {et~al.}(2015)\citenamefont {Faye},
  \citenamefont {Sahebsara},\ and\ \citenamefont
  {S\'en\'echal}}]{PhysRevB.92.085121}%
  \BibitemOpen
  \bibfield  {author} {\bibinfo {author} {\bibfnamefont {J.~P.~L.}\
  \bibnamefont {Faye}}, \bibinfo {author} {\bibfnamefont {P.}~\bibnamefont
  {Sahebsara}},\ and\ \bibinfo {author} {\bibfnamefont {D.}~\bibnamefont
  {S\'en\'echal}},\ }\bibfield  {title} {\bibinfo {title} {Chiral triplet
  superconductivity on the graphene lattice},\ }\href
  {https://doi.org/10.1103/PhysRevB.92.085121} {\bibfield  {journal} {\bibinfo
  {journal} {Phys. Rev. B}\ }\textbf {\bibinfo {volume} {92}},\ \bibinfo
  {pages} {085121} (\bibinfo {year} {2015})}\BibitemShut {NoStop}%
\bibitem [{\citenamefont {Kaba}\ and\ \citenamefont
  {S\'en\'echal}(2019)}]{PhysRevB.100.214507}%
  \BibitemOpen
  \bibfield  {author} {\bibinfo {author} {\bibfnamefont {S.-O.}\ \bibnamefont
  {Kaba}}\ and\ \bibinfo {author} {\bibfnamefont {D.}~\bibnamefont
  {S\'en\'echal}},\ }\bibfield  {title} {\bibinfo {title} {Group-theoretical
  classification of superconducting states of strontium ruthenate},\ }\href
  {https://doi.org/10.1103/PhysRevB.100.214507} {\bibfield  {journal} {\bibinfo
   {journal} {Phys. Rev. B}\ }\textbf {\bibinfo {volume} {100}},\ \bibinfo
  {pages} {214507} (\bibinfo {year} {2019})}\BibitemShut {NoStop}%
\end{thebibliography}
\providecommand{\noopsort}[1]{}\providecommand{\singleletter}[1]{#1}%

\end{document}